\documentclass[10pt,journal,compsoc]{IEEEtran}

\usepackage{cite}
\usepackage{amsmath,amssymb,amsfonts}
\usepackage{algorithmic}
\usepackage{graphicx}
\usepackage{textcomp}
\usepackage{xcolor}
\def\BibTeX{{\rm B\kern-.05em{\sc i\kern-.025em b}\kern-.08em
		T\kern-.1667em\lower.7ex\hbox{E}\kern-.125emX}}
\usepackage{subfigure}
\usepackage[english]{babel}
\usepackage{tabularx}
\usepackage{multirow}
\usepackage{array}
\usepackage{ifpdf}
\usepackage{url}
\usepackage{algorithm}
\usepackage{mdwlist}  
\usepackage{caption}
\usepackage{tikz}
\usepackage{diagbox}
\usepackage{hyperref}
\usepackage{listings}
\usepackage{enumitem}
\usepackage{wrapfig,lipsum,booktabs}
\usepackage{ragged2e}  

\newcommand{\ignore}[1]{}
\newcommand{\tabincell}[2]{\begin{tabular}{@{}#1@{}}#2\end{tabular}}
\newtheorem{insight}{Insight}
\newcommand*{\circled}[1]{\lower.7ex\hbox{\tikz\draw (0pt, 0pt)%
		circle (.45em) node {\makebox[1em][c]{\small #1}};}}
\newcommand{\D}{{\ensuremath{\sf D}}}
\newcommand{\defender}{{\ensuremath{\mathcal D}}}
\newcommand{\attacker}{{\ensuremath{\mathcal A}}}

\begin{document}
%
\title{Towards Making Deep Learning-based Vulnerability Detectors Robust}
%
%
%
%

\author{Zhen~Li, 
        Jing~Tang,
        Deqing~Zou,
        Qian~Chen,
        Shouhuai~Xu,
        Chao~Zhang,
        Yichen~Li,
        and Hai~Jin
        
\thanks{Corresponding authors: D. Zou and C. Zhang.}
\IEEEcompsocitemizethanks{
\IEEEcompsocthanksitem Z. Li is with the National Engineering Research Center for Big Data Technology and System, Services Computing Technology and System Lab, Cluster and Grid Computing Lab, Big Data Security Engineering Research Center, School of Cyber Science and Engineering, Huazhong University of Science and Technology, Wuhan 430074, China, and also with the University of Texas at San Antonio, San Antonio 78249, Texas, USA.
E-mail: lizhenhbu@gmail.com.

\IEEEcompsocthanksitem J. Tang and D. Zou are with the National Engineering Research Center for Big Data Technology and System, Services Computing Technology and System Lab, Cluster and Grid Computing Lab, Big Data Security Engineering Research Center, School of Cyber Science and Engineering, Huazhong University of Science and Technology, Wuhan 430074, China. 
E-mails: \{jingt, deqingzou\}@hust.edu.cn.

\IEEEcompsocthanksitem Q. Chen is with the University of Texas at San Antonio, San Antonio 78249, Texas, USA. 
Email: guenevereqian.chen@utsa.edu.

\IEEEcompsocthanksitem S. Xu is with the University of Colorado Colorado Springs, Colorado Springs 80918, Colorado, USA. This work was partly done at the University of Texas at San Antonio. 
E-mail: sxu@uccs.edu.

\IEEEcompsocthanksitem C. Zhang is with Tsinghua University, Beijing 100084, China. 
E-mail: chaoz@tsinghua.edu.cn.

\IEEEcompsocthanksitem Y. Li is with School of Cyber Science and Engineering, Huazhong University of Science and Technology, Wuhan 430074, China. 
E-mail: u201714846@hust.edu.cn. 

\IEEEcompsocthanksitem H. Jin is with the National Engineering Research Center for Big Data Technology and System, Services Computing Technology and System Lab, Cluster and Grid Computing Lab, Big Data Security Engineering Research Center, School of Computer Science and Technology, Huazhong University of Science and Technology, Wuhan 430074, China.
E-mail:  hjin@hust.edu.cn.
}

}

\IEEEtitleabstractindextext{%
\begin{abstract}
\justifying
Automatically detecting software vulnerabilities in source code is an important problem that has attracted much attention. In particular, deep learning-based vulnerability detectors, or DL-based detectors, are attractive because they do not need human experts to define features or patterns of vulnerabilities.
However, such detectors' robustness is unclear.
In this paper, we initiate the study in this aspect by demonstrating that DL-based detectors are not robust against simple code transformations, dubbed {\em attacks} in this paper, as these transformations may be leveraged for malicious purposes. As a first step towards making DL-based detectors robust against such attacks, we propose an innovative framework, dubbed {\em ZigZag}, which is centered at (i) decoupling {\em feature learning} and {\em classifier learning} and (ii) using a ZigZag-style strategy to iteratively refine them until they converge to robust features and robust classifiers.
Experimental results show that the ZigZag framework can substantially improve the robustness of DL-based detectors.
\end{abstract}

}

\maketitle

\IEEEdisplaynontitleabstractindextext

%
\IEEEpeerreviewmaketitle

\IEEEraisesectionheading{\section{Introduction}\label{sec:introduction}}
\IEEEPARstart{T}{he} problem of detecting software vulnerabilities is yet to be solved, as evidenced by a large number of vulnerabilities reported in the {\em Common Vulnerabilities and Exposures} (CVE) \cite{CVE}. 
The wide reuse of open-source software and the increasing complexity of software supply-chains make the problem even more imperative 
\cite{LinuxFoundation2020}. 
This can be justified by the Heartbleed vulnerability 
\cite{Heartbleed} and the software supply chain attack on the open-source npm package \cite{npm}, highlighting the importance of detecting vulnerabilities in source code.
The importance of the problem has motivated many studies involving two categories: 
{\em static analysis} which analyzes the software's source code and {\em dynamic analysis} which executes software and observes its behavior (e.g., fuzzing \cite{manes2019art,DBLP:conf/sp/GanZQTLPC18,DBLP:conf/ccs/BohmePNR17,DBLP:conf/uss/IspoglouAMP20}). 
In this paper, we focus on static analysis-based vulnerability detectors.
Static analysis-based detectors may analyze the source code in three ways: code similarity-based \cite{kim2017vuddy, DBLP:conf/acsac/LiZXJQH16, jang2012redebug} vs. rule-based \cite{FlawFinder,RATS,DBLP:conf/acsac/ViegaBKM00,Checkmarx,Coverity, DBLP:conf/ndss/GensSDS18,DBLP:conf/dimva/ShastryYRS16} vs. machine learning-based  \cite{yamaguchi2012generalized,neuhaus2007predicting,grieco2016toward,yamaguchi2015automatic} approaches. 
A recent development in machine learning-based detection is to use {\em Deep Learning} (DL). 
DL-based detectors are attractive because they do not need human experts to define features or patterns to represent vulnerabilities while achieving high effectiveness
\cite{vuldeepecker,SySeVR,DBLP:journals/corr/abs-2001-02334,DBLP:conf/ccs/LinZLPX17,duan2019vulsniper,zhou2019devign,DBLP:journals/tii/LinZLPXVM18,DBLP:conf/ijcnn/NguyenLLNDMQP19,liu2020cd,DBLP:conf/icics/LinXZX19,DBLP:conf/sigsoft/Sonnekalb19,DBLP:conf/icmla/RussellKHLHOEM18,DBLP:journals/tifs/WangYTTHFFBW21}. 

However, the robustness of DL-based vulnerability detectors is unclear, which motivates our study.
Although DL models are known to suffer from adversarial examples in many domains, such as image processing \cite{goodfellow2018defense,DBLP:journals/tnn/YuanHZL19}, speech recognition \cite{DBLP:conf/icml/QinCCGR19}, malware detection \cite{li2020sok,DBLP:journals/tifs/LiL20}, program analysis \cite{rabin2019testing,DBLP:conf/aaai/Zhang20,DBLP:journals/corr/abs-1910-07517}, and code authorship attribution \cite{DBLP:conf/uss/QuiringMR19},  
it is unknown whether or not this robustness problem equally holds for vulnerability detectors. This is so because ``adversarial vulnerability examples'' must be compiled and executed while preserving semantics and vulnerabilities, a requirement having no counterpart in the aforementioned domains.

\smallskip
\noindent{\bf Our contributions}. 
We initiate study on the robustness of DL-based vulnerability detectors, by making three contributions.
First, to understand the robustness of these detectors, we leverage semantics-preserving {\em code transformation} techniques, dubbed ``attacks'', to show that four representative DL-based detectors suffer from adversarial vulnerability examples. These detectors are representative because they operate at different granularities, use different vector representations, and employ different neural networks. For instance, some attacks against a state-of-the-art DL-based detector \cite{SySeVR} can cause its false-positive rate to increase from 7.0\% to 19.9\% and false-negative rate to increase from 9.9\% to 68.1\%.

Second, to make DL-based detectors robust against adversarial vulnerability examples, we propose an innovative framework, dubbed ``ZigZag''. The key insight is to decouple {\em feature learning} and {\em classifier learning} and make the resulting features and classifiers robust against code transformations. Specifically, ZigZag iteratively employs two classifiers, which have different decision boundaries but offer similar prediction results. 
In each iteration, the feature learning phase aims to extract {\em robust} features, which characterize input examples well and thus lead to similar, if not the same, predictions by the two different classifiers. Then, the classifier learning phase aims to train two {\em robust} classifiers, which have largely discrepant decision boundaries and few classification errors. 
In other words, the classifier learning phase optimizes two classifiers by increasing the discrepancy between their decision boundaries, whereas the feature learning phase optimizes features by reducing the two classifiers' prediction discrepancy. This procedure is iterated as many times as needed, explaining the term ``ZigZag''. When the iterative process converges, we obtain features and classifiers robust against code transformations.
To show the effectiveness of ZigZag, we apply it to the aforementioned detector \cite{SySeVR}. Experimental results show that when compared with the original detector, the hardened detector's false-positive rate (8.4\%) and false-negative rate (19.2\%) are much lower than the original 19.9\% and 68.1\% incurred by adversarial examples, respectively.

Third, our experiments are based on a new dataset we collected, which might be of independent value. This dataset is derived from the {\em National Vulnerability Database} (NVD) \cite{NVD} and the {\em Software Assurance Reference Dataset} (SARD) \cite{SARD}. 
It contains 6,803 
programs and their variants,
leading to 50,562 vulnerable examples and 80,043 non-vulnerable examples at the function level. 
We have made the dataset and source code of ZigZag publicly available at
\url{https://github.com/ZigZagframework/zigzag\_framework}.

\smallskip
	\noindent{\bf Paper organization}.
	We analyze the robustness of existing DL-based detectors in Section \ref{sec:Background}.
	Then, we present the design of our framework ZigZag and evaluate its robustness in Section \ref{sec:Design} and 
	Section \ref{sec:experiments-and-results}, respectively.
	Further, Section \ref{sec:Limitations} discusses the limitations and future work, 
	and Section \ref{sec:Related_work} reviews the related prior work.
	In the end, Section \ref{sec:Conclusion} concludes the paper. 
	Table \ref{Table_main_symbols} summarizes the main notations.

	\begin{table}[!t]
		\caption{\small Main notations used in the paper}
			\vspace{-0.2cm}
		\label{Table_main_symbols}
		\centering
		\scriptsize
		\begin{tabular}{|p{.13\textwidth}<{\centering}|p{.32\textwidth}|}
			\hline
			Notation & Meaning  \\
			\hline
			$M$ & The set of all 
			semantics-preserving code transformations
			\\ \hline
			$M_\attacker$ & $M_\attacker\subseteq M$ is available to the attacker\\
			\hline
			$M_\defender, M_{\defender,i}$ & $M_\defender \subseteq M$ is available to the defender; $M_{\defender,i} \subseteq M_{\defender}$ 
			is a subset of transformations used in a specific experiment \\ \hline
			$P$ & $P=\{p_1, \ldots, p_n\}$ is a set of training programs \\
			\hline
			$P^+$ & A set of programs expanded from $P$ by including the programs that are transformed from the ones in $P$ via some code transformations in $M_\defender$\\
			\hline
			$\D$ & A DL-based detector learned from programs in $P$\\
			\hline
			$\D^+$ & A ZigZag-enabled detector $\D$ learned from programs in $P^+$\\
			\hline
			$q,q^+$ & Program $q$ has a vulnerability that can be detected by $\D$; $q^+$ is a semantics-preserving transformation of $q$ and has a vulnerability that cannot be detected by $\D$\\
			\hline
			$Q,Q^+$ & $Q$ is a set of original test programs; $Q^+$ is a set of target programs composed of the programs in $Q$ and their manipulated programs\\
			\hline
			$\Pr(\D,q)$ & The probability that $\D$ predicts program $q$ as vulnerable\\
			\hline
			$\delta$ & A threshold probability, whereby $\D$ predicts $q$ as vulnerable if $\Pr(\D,q)>\delta$ and non-vulnerable otherwise.\\
			\hline
			$X$ & $X=\{{\bf x_1}, \dots, {\bf x_\mu}\}$ is a set of vectors corresponding to all examples generated from programs in $P$,  where ${\bf x_s}$ ($1 \leq s \leq \mu$) is the vector of an example with label $y_s$\\
			\hline
			$X'$ & $X'=\{{\bf x'_1}, \dots, {\bf x'_\nu}\}$ is a set of all examples generated from the program in $P^+\!\!\!-P$,  where ${\bf x'_w}$ ($1 \leq w \leq {\bf \nu}$) is the vector of an example with label $y'_w$\\
			\hline
			$X''$ & $X''=\{{\bf x''_1}, \dots, {\bf x''_\gamma}\}$ is a set of hard examples (i.e., false positives and false negatives) in $X'$, where ${\bf x''_d}$ ($1 \leq d \leq \gamma\;$) is the vector of an example with  label $y''_d$\\
			\hline
			$F, F^\ast$ & Feature generators used in $\D^+$\\
			\hline
			$C_1, C_2, C_1^\ast, C_2^\ast$ & Classifiers used in $\D^+$\\
			\hline
			$c_1({\bf x_s}, F), c_2({\bf x_s}, F),$ $c_1^\ast({\bf x_s}, F), c_2^\ast({\bf x_s}, F)$ & The probability that $C_1$/ $C_2$/ $C_1^\ast$/ $C_2^\ast$ predicts an example ${\bf x_s}$ as vulnerable while using feature generator $F$\\
			\hline
		\end{tabular}
	\end{table}

\section{Robustness of DL-based Detectors}
\label{sec:Background}
\begin{figure}[!t]
	\centering
	\includegraphics[width=.4\textwidth]{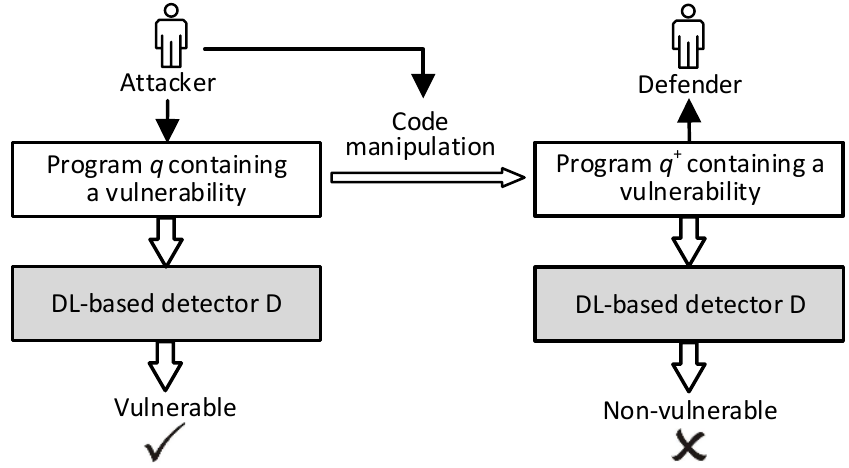}
	\vspace{-0.2cm}
	\caption{Illustration of an attack against DL-based detector $\D$ 
	}
	\label{Fig_attack_model}
	\vspace{-0.4cm}
\end{figure}

A DL-based detector $\D$ is trained from a set of programs in source code. The defender uses $\D$ to determine whether a given {\em target program} in source code is vulnerable or not. Fig. \ref{Fig_attack_model} highlights the basic idea:
the attacker attempts to manipulate a program, while preserving its semantics, to cause $\D$ to classify (i) a manipulated program containing no vulnerability as ``vulnerable'' or (ii) a manipulated program containing a vulnerability as ``non-vulnerable''. 
In this paper, we call manipulated programs  {\em adversarial examples} regardless of whether the degree of code manipulation is small or not.

\subsection{Attack Requirements}
\label{sec:vulnerability-backdoor-attack-requirements}

The {\em first} attack requirement is to preserve the semantics of a program. This is important because the manipulated program should be as useful as the original program. 
The {\em second} attack requirement is not to use obfuscation technique because users may not use any obfuscated code from a third party that is not known to be trustworthy (in fear of malicious code).
The {\em third} attack requirement is the preservation of the vulnerability itself. This means that given a piece of vulnerable code, where the vulnerability can be detected by some existing DL-based detectors, the manipulated code remains vulnerable but the vulnerability it contains can evade those detectors. This is important because the attacker's goal is to make vulnerabilities evade vulnerability detectors.

\subsection{Attack Experiments}
\label{subsec:attack_exp}

\subsubsection{Selecting DL-based Detectors} 
Since DL-based detectors can be characterized by the {\em granularity} (e.g., function \cite{DBLP:conf/icics/LinXZX19,duan2019vulsniper,zhou2019devign} vs. program slice \cite{vuldeepecker,SySeVR,DBLP:journals/corr/abs-2001-02334}), the {\em vector representation} (e.g., sequence-based \cite{vuldeepecker,SySeVR, DBLP:journals/corr/abs-2001-02334} vs. Abstract Syntax Tree or AST-based \cite{DBLP:conf/ccs/LinZLPX17, DBLP:journals/tii/LinZLPXVM18,liu2020cd}), and the {\em neural network} (e.g., Bidirectional Gated Recurrent Unit or BGRU \cite{SySeVR} vs. Bidirectional Long Short-Term Memory or BLSTM
\cite{vuldeepecker,DBLP:journals/tii/LinZLPXVM18} vs. Convolutional Neural Network or CNN \cite{DBLP:conf/icmla/RussellKHLHOEM18}), we consider the following four
DL-based detectors.

\noindent{\bf 
Program Slice + Sequence + BGRU}. It can be instantiated as the detector SySeVR \cite{SySeVR}, which is an extended version of VulDeePecker \cite{vuldeepecker},
is publicly available, and operates at the fine granularity in that each program is represented by multiple program slices.
A program slice is composed of a small number of program statements that are  
semantically related to each other. A slice is parsed as a sequence of tokens  (e.g., identifiers, operators, constants, and keywords) and transformed into a vector. This vector representation is used to train a BGRU model for classifying program slices as vulnerable or not. 
	
\noindent{\bf 
Function + Sequence + CNN}. 
It can be instantiated as the detector presented in \cite{DBLP:conf/icmla/RussellKHLHOEM18}, which operates at the coarse granularity in that each function is treated as a unit. Specifically, each program is divided into multiple functions; each function is interpreted as a sequence of tokens
and transformed into a vector; and these vectors are used for training a CNN model, which classifies the functions as vulnerable or not.

\noindent{\bf 
Function + Sequence + BLSTM}. It can be instantiated as the detector presented in \cite{DBLP:conf/icics/LinXZX19}. It operates at a coarse granularity by dividing a program into multiple functions, representing each function as a sequence of tokens, and training a BLSTM model to classify functions as vulnerable or not. 

\noindent{\bf Function + AST + BLSTM}. It can be instantiated as the detector that extends and enhances the DL-based detector \cite{DBLP:conf/icics/LinXZX19} to capture more syntactic and structural information of program source code, by replacing its sequence-based representation 
with AST-based representation \cite{DBLP:conf/sigsoft/Sonnekalb19} and using the {\em code2vec} tool \cite{alon2019code2vec,DBLP:conf/msr/KovalenkoBBB19} to aggregate multiple AST paths into a vector.

\subsubsection{Preparing Dataset}
The present study needs a dataset that satisfies the following requirements: (i) the dataset can support the generation of examples at different granularities;
(ii) the programs in the dataset can be compiled for code transformation purposes; and (iii) the dataset should contain vulnerabilities in real-world software for training because our goal is to detect real-world software vulnerabilities which may be different from synthetic vulnerabilities. 
Because existing datasets \cite{DBLP:conf/icics/LinXZX19,zhou2019devign,DBLP:conf/icmla/RussellKHLHOEM18,SySeVR,vuldeepecker,DBLP:conf/nips/HarerOLRRKC18} do not satisfy the preceding three requirements,
we create a new dataset by considering two vulnerability sources:
NVD \cite{NVD} and SARD \cite{SARD}. 
For NVD, we collect: (i) vulnerable program files that are reported before 2017
and belong to open-source software written in C; and (ii) their patches, which
can be obtained from the software vendors' websites. The rationale for (i) 
is that we conduct experiments on real-world open-source software to detect vulnerabilities reported from 2017 to 2019 
(Section \ref{sec:attack-experiments-results} and Section \ref{subsec:experiments_RQ1}). 
For SARD, each program is labeled as {\em good} (not vulnerable), {\em bad} (vulnerable), or
{\em mixed} (vulnerable functions and their patched versions).
In total, we collect 6,803 programs, 
each of which is vulnerable or patched. 
We take vulnerable (i.e., positive) examples and
non-vulnerable (i.e., negative) examples at the function level as the ground truth, because each vulnerability can map to a function and each function has at most one vulnerability in our dataset.
The 6,803 original programs includes 6,865 vulnerable examples and 10,843 non-vulnerable examples.
The 6,803 programs and their variants generated by applying code transformations include 50,562 vulnerable examples and 80,043 non-vulnerable examples in total.

\begin{figure*}[!htb]
	\centering
	\includegraphics[width=.9\textwidth]{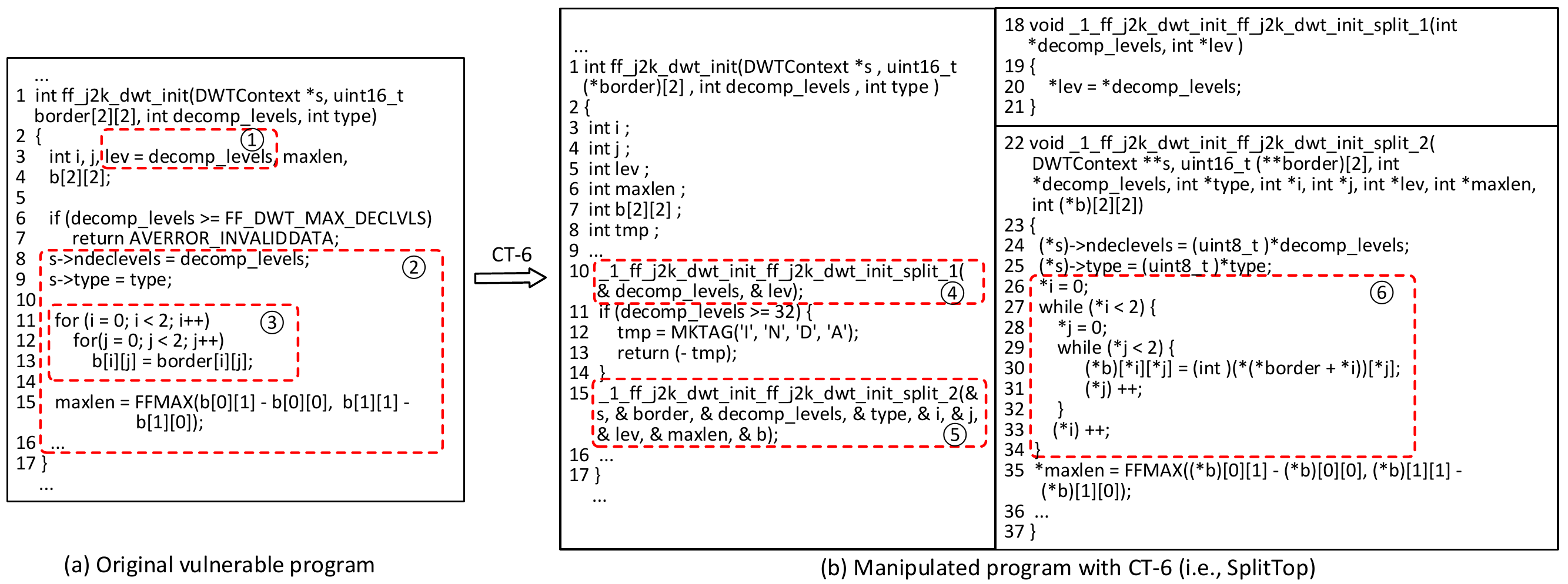}
	\vspace{-0.2cm}
\caption{A vulnerable program (CVE-2012-0849) and its manipulated version obtained by applying the code transformation CT-6
	}
	\label{Fig_CVE_example}
\end{figure*}

\subsubsection{Attack Methods} To demonstrate the feasibility of the attacks, we leverage real-world code transformation tools to launch attacks because they are designed to preserve program semantics. 
There are multiple real-world code transformation tools \cite{Tigress,Stunnix,Sourceformatx,Coccinelle};
we choose Tigress \cite{Tigress} 
because it provides various code transformations without obfuscating code.
Table \ref{Table_code_variations} describes 8 code transformations, denoted by $\text{CT-1}, \ldots, \text{CT-8}$, 
which are selected from what are offered by Tigress.
We apply each of the 8 code transformations to each of the original programs to generate manipulated programs. 
Fig. \ref{Fig_CVE_example}(a) illustrates a vulnerable program containing an integer overflow vulnerability CVE-2012-0849 
(vulnerable Line 6). 
Fig. \ref{Fig_CVE_example}(b) shows its manipulated version obtained by applying the code transformation CT-6 (i.e., SplitTop). 
CT-6 splits the original vulnerable function  {\em ff\_j2k\_dwt\_init} into two new functions {\em \_1\_ff\_j2k\_dwt\_init\_ff\_j2k\_dwt\_init\_split\_1} and {\em \_1\_ff\_j2k\_dwt\_init\_ff\_j2k\_dwt\_init\_split\_2}.
The code in the dashed box highlighted with \circled{1} in Fig.  \ref{Fig_CVE_example}(a) corresponds to the code in the dashed box highlighted with \circled{4} in Fig. \ref{Fig_CVE_example}(b); the code in the dashed box highlighted with \circled{2} in Fig.  \ref{Fig_CVE_example}(a) corresponds to the code in the dashed box highlighted with \circled{5} in Fig.  \ref{Fig_CVE_example}(b). 
CT-6 also replaces the {\tt for} loop with the {\tt while} loop (i.e., the code in the dashed box highlighted with \circled{3} in Fig.  \ref{Fig_CVE_example}(a) and the code in the dashed box highlighted with \circled{6} in Fig.  \ref{Fig_CVE_example}(b)), replaces arrays with pointers (e.g., Line 15 in Fig.  \ref{Fig_CVE_example}(a) and Line 35 in Fig.  \ref{Fig_CVE_example}(b)), and replaces macro definition identifiers with static values (e.g., Line 6 in Fig.  \ref{Fig_CVE_example}(a) and Line 11 in Fig.  \ref{Fig_CVE_example}(b)).

\begin{table}[!t]
	\caption{The 8 code transformations we use for attacks
	}
	\label{Table_code_variations}
	\centering
	\scriptsize
	\begin{tabular}{|c|p{.08\textwidth}<{\centering}|p{.3\textwidth}|}
		\hline
		No. & Name & Description \\
		\hline
		CT-1 & EncodeStrings & Replace the literal strings with calls to the function that generates these literal strings. \\
		\hline
		CT-2 & RndArgs & Reorder function arguments and/or add bogus arguments. \\
		\hline
		CT-3 & Flatten & Remove some control flows from a function. 
		\\
		\hline
		CT-4 & MergeSimple & Merge multiple functions into one without control-flow flattening. 
		\\
		\hline
		CT-5 & MergeFlatten & Merge multiple functions into one with control-flow flattening. 
		\\
		\hline
		CT-6 & SplitTop & Split top-level statements into multiple functions. \\
		\hline
		CT-7 & SplitBlock & Split a basic block into multiple functions.\\
		\hline
		CT-8 & SplitRecursive & Split a basic block into multiple functions, and split the calls to split functions.\\
		\hline
	\end{tabular}
	\vspace{-0.6cm}
\end{table}

\subsubsection{Experimental Results}
\label{sec:attack-experiments-results}
Our attacks satisfy the aforementioned attack requirements as follows. The requirement of {\em semantics preservation} is assured by Tigress, which is used to conduct code transformations and is designed to preserve program semantics. The requirement of {\em no-obfuscation} is satisfied by choosing 8 code transformations that do not involve any obfuscation operations.
The requirement of {\em vulnerability preservation} is assured by manual examination. 
To check whether semantics-preserving transformations can preserve vulnerabilities, we add a flag to each vulnerable line of code in the original program to trace 
its corresponding line(s) of code in the manipulated program.
We randomly select 200 vulnerable programs and manually check if the manipulated programs are still vulnerable. It takes about 105 hours of domain experts to confirm that the manipulated programs contain the same vulnerabilities as in the original programs.

\begin{table*}[!t]
\caption{Experimental results showing that the 4 DL-based detectors are lack of robustness against code transformations (metrics unit: \%) 
}
	\vspace{-0.2cm}
	\label{Table_original_model_slice}
	\scriptsize
	\centering
	\begin{tabular}{|c|p{.058\textwidth}<{\centering}|p{.058\textwidth}<{\centering}|p{.058\textwidth}<{\centering}||p{.045\textwidth}<{\centering}|p{.045\textwidth}<{\centering}|p{.045\textwidth}<{\centering}||p{.05\textwidth}<{\centering}|p{.05\textwidth}<{\centering}|p{.05\textwidth}<{\centering}||p{.04\textwidth}<{\centering}|p{.04\textwidth}<{\centering}|p{.04\textwidth}<{\centering}|}
		\hline
		& \multicolumn{3}{c||}{Program Slice + Sequence + BGRU} & \multicolumn{3}{c||}{Function + Sequence + CNN} & \multicolumn{3}{c||}{Function + Sequence + BLSTM} & \multicolumn{3}{c|}{Function + AST + BLSTM}\\
		\hline
		CT & FPR & FNR & F1 & FPR & FNR & F1 & FPR & FNR & F1 & FPR & FNR & F1\\
		\hline
		n/a & 7.0 & 9.9 & 88.1 & 10.5 & 18.8 & 82.1 & 7.3 & 12.0 & 88.2 & 7.3 & 12.4 & 88.0 \\
		\hline
		CT-1 & 15.5 & 66.4 & 39.4 & 38.8 & 45.7 & 50.7 & 35.0 & 40.4 & 55.8 & 34.3 & 40.3 & 56.1 \\
		\hline
		CT-2 & 16.7 & 67.9 & 37.3 & 24.8 & 39.4 & 61.2 & 21.8 & 35.2 & 65.5 & 19.9 & 34.5 & 67.0 \\
		\hline
		CT-3 & 22.9 & 75.4 & 27.8 & 44.5 & 47.3 & 47.8 & 40.3 & 43.5 & 51.8 & 39.2 & 43.5 & 52.2 \\
		\hline
		CT-4 & 24.0 & 77.7 & 25.2 & 43.6 & 67.0 & 34.8 & 41.0 & 64.9 & 37.3 & 38.5 & 64.9 & 37.9 \\
		\hline
		CT-5 & 24.0 & 78.6 & 24.2 & 42.6 & 70.4 & 32.0 & 40.8 & 68.2 & 34.4 & 40.2 & 67.9 & 34.8 \\
		\hline
		CT-6 & 23.9 & 77.5 & 25.4 & 58.3 & 57.6 & 35.9 & 55.3 & 54.7 & 38.7 & 54.1 & 53.8 & 39.6 \\
		\hline
		CT-7 & 23.7 & 76.8 & 26.2 & 59.2 & 60.3 & 34.5 & 57.1 & 57.7 & 36.9 & 55.9 & 56.7 & 38.0 \\
		\hline
		CT-8 & 24.0 & 77.8 & 25.1 & 56.5 & 59.7 & 35.6 & 54.8 & 58.3 & 37.0 & 54.0 & 57.1 & 38.1 \\
		\hline
		\hline
		Total & 19.9 & 68.1 & 35.7 & 41.6 & 49.9 & 47.0 & 38.7 & 46.3 & 50.6 & 37.7 & 45.9 & 51.3 \\
		\hline
	\end{tabular}
	\vspace{-0.4cm}
\end{table*}

\smallskip
\noindent{\bf Evaluation of vulnerability detection evasion}.
To show the lack of robustness of DL-based detectors,  
we conduct attacks against 
four DL-based detectors. We randomly choose 80\% of the 6,803 programs 
for training and use the rest
for test. Target programs $Q^+$ involve the original test programs $Q$ and their manipulated programs with 8 code transformations, which are available to the attacker. At the function level, the training programs contain 4,079 vulnerable examples and 6,530 non-vulnerable examples; the target programs contain 17,516 vulnerable examples and 28,206 non-vulnerable examples. 

Let {\sf TP} denote true positives, {\sf FP} denote false positives, {\sf TN} denote true negatives, and {\sf FN} denote false negatives.
We use three standard metrics for evaluation \cite{DBLP:journals/csur/PendletonGCX17}:
(i) False-Positive Rate $\text{FPR}=\frac{{\sf FP}}{{\sf FP}+{\sf TN}}$; 
(ii) False-Negative Rate $\text{FNR}=\frac{{\sf FN}}{{\sf TP}+{\sf FN}}$; 
(iii) overall effectiveness or $\text{F1}$-measure $\text{F1}=\frac{2 \cdot \text{P} \cdot (1-\text{FNR})}{\text{P}+(1-\text{FNR})}$, where precision $\text{P}=\frac{{\sf TP}}{{\sf TP}+{\sf FP}}$.  
We train four DL-based detectors and choose the hyperparameters that lead to the highest F1.
Table \ref{Table_original_model_slice} summarizes the results. 
Compared with the function-level detectors, the program slice-level detector achieves better results for original test programs $Q$ with a 1.4\% lower FPR, a 4.5\% lower FNR, and a 2.0\% higher F1 on average, but achieves a 19.4\% lower FPR, a 20.7\% higher FNR and a 13.9\% lower F1 for target programs $Q^+$ on average.
This indicates that the program slice-level detector misses many more vulnerabilities in manipulated programs. We speculate that this is caused by the following:
a program slice has a finer granularity than a function, thus the detector at the program slice-granularity is more sensitive to the changes of vulnerable code. 
In contrast, the coarser-grained function-level detector can accommodate more changes in both vulnerable and non-vulnerable code. 
Therefore, the function-level detector causes more false-positives and fewer false-negatives for manipulated programs. 
In addition, each detector exhibits similar phenomena with respect to different code transformations. Take the ``Program Slice + Sequence + BGRU'' detector for instance. We observe that the manipulated programs achieve a high FPR of 21.8\%, a high FNR of 74.8\%, and a low F1 of 28.8\% on average, which indicates that the DL-based detectors can  easily make mistakes by manipulating programs.

\begin{table}[!t]
\caption{
The vulnerabilities in the three software products that evade the DL-based detectors}

	\vspace{-0.2cm}
	\label{Table_open_source_software_CVE}
	\centering
	\tiny
	\begin{tabular}{|p{.065\textwidth}<{\centering}|c|c|p{.15\textwidth}<{\centering}|}
		\hline
		DL-based detector & Software product & CVE ID & Code transformations\\
		\hline
		{\multirow{18}{*}{\tabincell{c}{Program\\ Slice+\\Sequence+ \\BGRU}}} & FFmpeg 2.8.2 & CVE-2017-9608 & CT-3, CT-5, CT-6 \\
		\cline{2-4}
		& FFmpeg 2.8.2 & CVE-2018-14395 & CT-3, CT-5, CT-6 \\
		\cline{2-4}
		& FFmpeg 2.8.2 & CVE-2018-14394 & CT-3, CT-4, CT-5 \\
		\cline{2-4}
		& FFmpeg 2.8.2 & CVE-2018-1999010 & CT-1, CT-3, CT-4, CT-5, CT-6, CT-7 \\
		\cline{2-4}
		& FFmpeg 2.8.2 & CVE-2019-12730 & CT-1, CT-2 \\
		\cline{2-4}
		& Wireshark 2.0.5 & CVE-2017-6467 & CT-1, CT-2, CT-3 \\
		\cline{2-4}
		& Wireshark 2.0.5 & CVE-2017-6468 & CT-1, CT-2, CT-3, CT-5, CT-6 \\
		\cline{2-4}
		& Wireshark 2.0.5 & CVE-2017-6469 & CT-2, CT-3, CT-4, CT-5, CT-8 \\
		\cline{2-4}
		& Wireshark 2.0.5 & CVE-2017-6474 & CT-4, CT-5, CT-6 \\
		\cline{2-4}
		& Wireshark 2.0.5 & CVE-2017-7702 & CT-1, CT-2 \\
		\cline{2-4}
		& Wireshark 2.0.5 & CVE-2017-9345 & CT-3, CT-4, CT-6, CT-7, CT-8 \\
		\cline{2-4}
 	    & Wireshark 2.0.5 & CVE-2017-11410 & CT-4, CT-6, CT-7, CT-8 \\
		\cline{2-4}
		& Wireshark 2.0.5 & CVE-2017-11411 & CT-4, CT-6, CT-7, CT-8 \\
		\cline{2-4}
		& Wireshark 2.0.5 & CVE-2017-13767 & CT-1, CT-3, CT-4, CT-7 \\
		\cline{2-4}
		& OpenSSL 1.1.0 & CVE-2017-3730 & CT-1, CT-2, CT-6, CT-7 \\
		\cline{2-4}
		& OpenSSL 1.1.0 & CVE-2017-3733 & CT-1, CT-2, CT-3, CT-4 \\
		\cline{2-4}
		& OpenSSL 1.1.0 & CVE-2018-0732 & CT-3, CT-4, CT-5, CT-6, CT-7 \\
		\cline{2-4}
		& OpenSSL 1.1.0 & CVE-2019-1543 & CT-1, CT-2, CT-3, CT-4 \\
		\cline{2-4}
		& OpenSSL 1.1.0 & CVE-2019-1563 & CT-1, CT-2, CT-3, CT-4 \\
		\hline
		{\multirow{8}{*}{\tabincell{c}{Function+\\Sequence+ \\CNN}}} & FFmpeg 2.8.2 & CVE-2018-9996 & CT-1, CT-3 \\
		\cline{2-4}
		& FFmpeg 2.8.2 & CVE-2018-14395 & CT-3, CT-4 \\
		\cline{2-4}
		& FFmpeg 2.8.2 & CVE-2018-1999010 & CT-1, CT-3, CT-5 \\
		\cline{2-4}
		& Wireshark 2.0.5 & CVE-2017-6467 & CT-2, CT-3 \\
		\cline{2-4}
		& Wireshark 2.0.5 & CVE-2017-6474 & CT-4, CT-5, CT-6 \\
		\cline{2-4}
		& Wireshark 2.0.5 & CVE-2017-9344 & CT-6, CT-7 \\
		\cline{2-4}
		& OpenSSL 1.1.0 & CVE-2017-3733 & CT-1, CT-2 \\
		\cline{2-4}
		& OpenSSL 1.1.0 & CVE-2018-0735 & CT-4, CT-5 \\
		\hline
		{\multirow{10}{*}{\tabincell{c}{Function+\\Sequence+ \\BLSTM}}} & FFmpeg 2.8.2 & CVE-2017-9608 & CT-3, CT-5 \\
		\cline{2-4}
		& FFmpeg 2.8.2 & CVE-2018-14394 & CT-3, CT-4, CT-5 \\
		\cline{2-4}
		& FFmpeg 2.8.2 & CVE-2018-1999010 & CT-1, CT-3, CT-5 \\
		\cline{2-4}
		& Wireshark 2.0.5 & CVE-2017-6467 & CT-2, CT-3 \\
		\cline{2-4}
		& Wireshark 2.0.5 & CVE=2017-6474 & CT-4, CT-5, CT-6 \\
		\cline{2-4}
		& Wireshark 2.0.5 & CVE-2017-9345 & CT-3, CT-5, CT-6 \\
		\cline{2-4}
		& Wireshark 2.0.5 & CVE-2017-11411 & CT-4, CT-5, CT-6 \\
		\cline{2-4}
		& OpenSSL 1.1.0 & CVE-2017-3733 & CT-1, CT-2, CT-3 \\
		\cline{2-4}
		& OpenSSL 1.1.0 & CVE-2018-0735 & CT-3, CT-4, CT-5 \\
		\cline{2-4}
		& OpenSSL 1.1.0 & CVE-2018-0737 & CT-3, CT-4, CT-5 \\
		\hline
		{\multirow{12}{*}{\tabincell{c}{Function+\\AST+ \\BLSTM}}} & FFmpeg 2.8.2 & CVE-2017-9608 & CT-3, CT-5, CT-6 \\
		\cline{2-4}
		& FFmpeg 2.8.2 & CVE-2018-14394 & CT-3, CT-4, CT-5 \\
		\cline{2-4}
		& FFmpeg 2.8.2 & CVE-2017-9996 & CT-1, CT-3, CT-5 \\
		\cline{2-4}
		& FFmpeg 2.8.2 & CVE-2019-12730 & CT-1, CT-2 \\
		\cline{2-4}
		& Wireshark 2.0.5 & CVE-2017-6468 & CT-2, CT-3 \\
		\cline{2-4}
		& Wireshark 2.0.5 & CVE-2017-6474 & CT-4, CT-5, CT-6 \\
		\cline{2-4}
		& Wireshark 2.0.5 & CVE-2017-9345 & CT-3, CT-4, CT-6, CT-7, CT-8 \\
		\cline{2-4}
		& Wireshark 2.0.5 & CVE-2017-11410 & CT-4, CT-6, CT-7, CT-8 \\
		\cline{2-4}
		& Wireshark 2.0.5 & CVE-2017-13766 & CT-3, CT-4, CT-6 \\
		\cline{2-4}
		& OpenSSL 1.1.0 & CVE-2017-3733 & CT-1, CT-2, CT-3 \\
		\cline{2-4}
		& OpenSSL 1.1.0 & CVE-2018-0734 & CT-1, CT-2, CT-3 \\
		\cline{2-4}
		& OpenSSL 1.1.0 & CVE-2019-1543 & CT-1, CT-2, CT-3,CT-4 \\
		\hline
	\end{tabular}
	\vspace{-0.6cm}
\end{table}

To show the feasibility of the attack against real-world open-source software, we test it against three open-source software products 
to detect the vulnerabilities reported in the NVD from 2017 to 2019, while recalling that these detectors are trained using the vulnerabilities reported prior to 2017. 
We use the four DL-based detectors to detect vulnerabilities in three software products and their manipulated versions. We observe that the program slice-level detectors can detect more vulnerabilities than the function-level detectors and some vulnerabilities detected by different detectors are the same.
Table \ref{Table_open_source_software_CVE} summarizes the vulnerabilities in three software products that can evade the DL-based detectors. 
We observe that there are 19 vulnerabilities, 8 vulnerabilities, 10 vulnerabilities, and 12 vulnerabilities that can respectively evade the ``Program Slice + Sequence + BGRU'', the ``Function + Sequence + CNN'', the ``Function + Sequence + BLSTM'', and the ``Function + AST + BLSTM'' detectors.
Considering the ``Program Slice + Sequence + BGRU'' detector, we observe that 5 vulnerabilities in FFmpeg 2.8.2, 9 vulnerabilities in Wireshark 2.0.5, and 5 vulnerabilities in OpenSSL 1.1.0 are missed; 
these vulnerabilities are listed in Table \ref{Table_open_source_software_CVE}.
In summary, 

\begin{insight}
{\em DL-based vulnerability detectors are not robust against evasion.}
\end{insight}

\section{ZigZag Framework}
\label{sec:Design}

\subsection{Characterizing DL-based Detectors}

\begin{figure*}[!t]
	\centering
	\includegraphics[width=.98\textwidth]{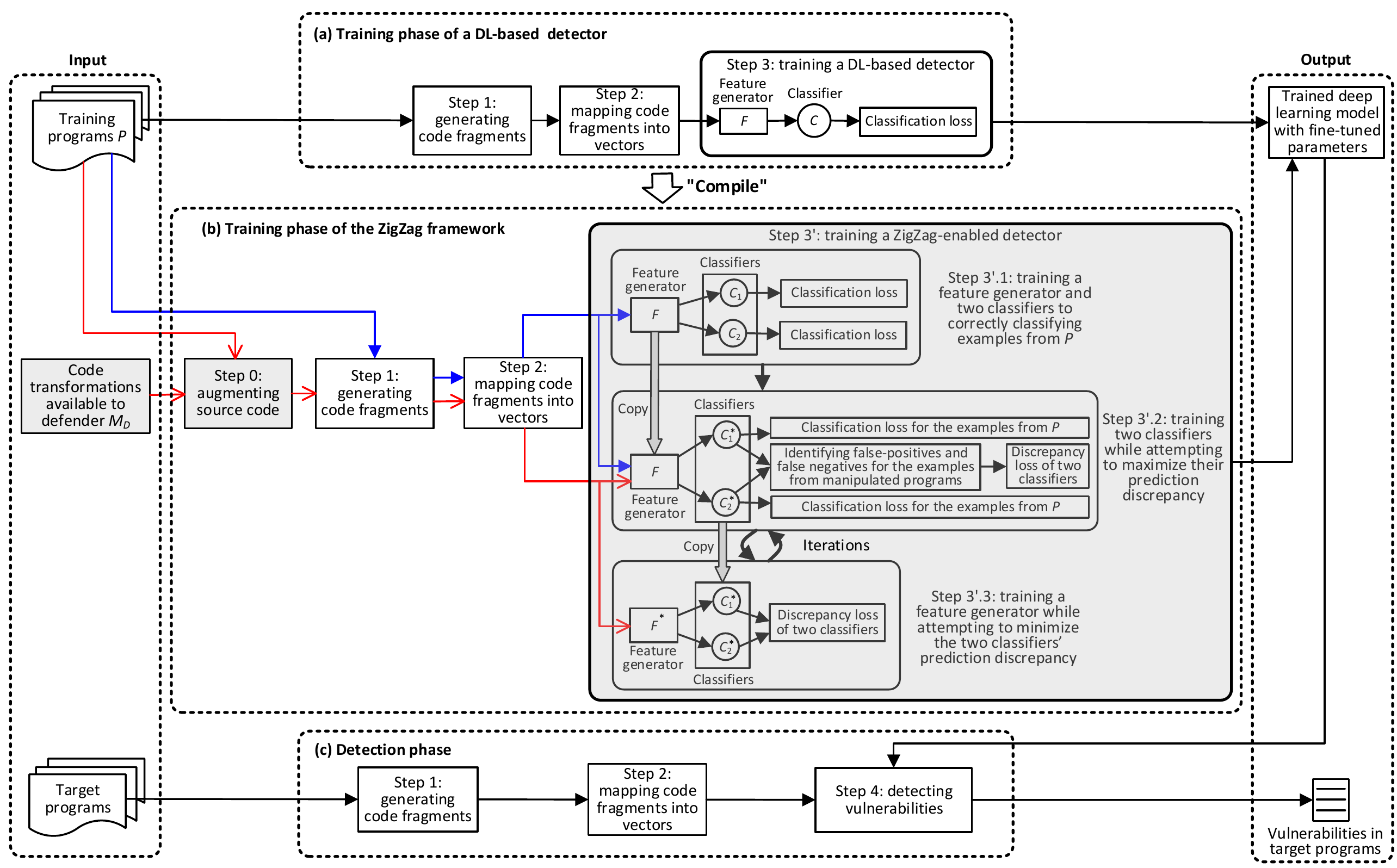}
	\vspace{-0.2cm}
	\caption{Overview of the ZigZag framework.  
	The ZigZag framework attempts to ``compile'' a DL-based detector into a new vulnerability detector robust against code transformations. 
	The new steps (i.e., Steps 0 and 3') introduced by the ZigZag framework are highlighted with shaded boxes. 
	Since the two data flows share Steps 1 and 2, we use blue arrows and red arrows to distinguish the inputs to Step 3'.}
	\label{Fig_Overview_of_our_method}
	\vspace{-0.6cm}
\end{figure*}

Fig. \ref{Fig_Overview_of_our_method}(a) and (c) depict the training phase and detection phase of a DL-based detector. The training phase consists of Steps 1, 2, and 3; the detection phase consists of Steps 1, 2, and 4. These steps are elaborated below.

\noindent{\bf Step 1: generating code fragments}. 
In the training phase, {\em training programs} are used to train a DL-based detector.
In the detection phase, the DL-based detector is used to detect whether the {\em target programs} contain vulnerabilities or not. 
A detector operates on code fragments at a certain granularity, such as function  \cite{DBLP:conf/icics/LinXZX19,duan2019vulsniper,zhou2019devign} 
and program slice \cite{vuldeepecker, SySeVR,DBLP:journals/corr/abs-2001-02334}.  
This step extracts code fragments from each training program and each target program at the desired granularity. A code fragment extracted from a training program is labeled as vulnerable if it contains  vulnerable statements and labeled as non-vulnerable otherwise.

\noindent{\bf Step 2: mapping code fragments to vectors}. This step maps each code fragment into an appropriate form of representation (e.g., a sequence of tokens \cite{vuldeepecker,SySeVR, DBLP:journals/corr/abs-2001-02334} or an abstract syntax tree \cite{DBLP:conf/ccs/LinZLPX17, DBLP:journals/tii/LinZLPXVM18,liu2020cd,DBLP:conf/sigsoft/Sonnekalb19}), depending on the specifics of DL-based detectors. 
The code representation is then embedded into a vector by, for example, the concatenation of the vectors corresponding to the tokens.  

\noindent{\bf Step 3: training a DL-based detector}. This step only applies to the training phase. It uses the vectors corresponding to the code fragments 
and their labels to train a deep neural network, such as BGRU \cite{SySeVR}, BLSTM \cite{vuldeepecker,DBLP:journals/tii/LinZLPXVM18}, or CNN \cite{DBLP:conf/icmla/RussellKHLHOEM18}. 

\noindent{\bf Step 4: detecting vulnerabilities}. This step only applies to the detection phase. It uses the trained DL-based detector to determine whether a vector, which corresponds to a code fragment extracted from a target program, is vulnerable or not. 

\subsection{System and Threat Model}
\label{subsec:threat_model}
In the {\em system} model, we consider a defender, denoted $\defender$. The defender trains a DL-based detector, denoted by $\D$, from a set of training programs $P=\{p_1, \ldots, p_n\}$. Let $M$ be the set of all possible code transformation methods whereby one can modify or manipulate one program into another program such that these two programs have the same semantics, dubbed {\em semantics-preserving code transformations}. 
Let $\Pr(\D,q)$ denote the probability that program $q$ is vulnerable, according to detector $\D$. 
For the given threshold probability $\delta$, $\D$ predicts $q$ as vulnerable if $\Pr(\D,q)>\delta$ and non-vulnerable otherwise.

In the {\em threat} model, the attacker, denoted by $\attacker$, has access to a set $M_\attacker$, where $M_{\attacker} \subseteq M$, of code transformation methods by which the attacker can manipulate a program $q$ into a new program $q^+$ via  semantics-preserving code transformations. The attacker's objective is to induce mistakes from the defender's detector $\D$, namely achieving:
\begin{eqnarray}
\Pr(\D,q^+)\left\{
\begin{array}{ll}
>\delta & \text{if}~ \Pr(\D,q)\leq \delta;\label{eq:attack-casusing-FP}\\
\leq\delta & \text{if}~ \Pr(\D,q)> \delta,\label{eq:attack-casusing-FN}
\end{array}
\right.
\end{eqnarray}
where 
$\Pr(\D,q^+)> \delta$ means a false positive and
$\Pr(\D,q^+)\leq \delta$ means a false negative. $\attacker$ may be interested in causing false negatives, which is known as the {\em evasion} attack  \cite{DBLP:conf/ccs/HuangJNRT11}.

The {\em design objective} is to harden $\D$ into $\D^+$ such that $\D^+$ can 
detect the vulnerability in $q^+$, namely achieving:
\begin{eqnarray}
\Pr(\D^+,q^+)\left\{
\begin{array}{ll}
\leq \delta & \text{if}~ \Pr(\D,q)\leq \delta;\\
>\delta & \text{if}~ \Pr(\D,q)> \delta.
\end{array}
\right.
\label{eq:design-objective}
\end{eqnarray}

\subsection{The ZigZag Framework}
To achieve the design objective, 
it is intuitive to allow the defender to extend the set $P$ of training programs into a new set $P^+$ by mimicking 
what the attacker would do to evade $\D$. This enhanced set $P^+$ is leveraged to harden $\D$ into $\D^+$.
To produce $P^+$, the defender needs to use some semantics-preserving code transformation methods. Let $M_\defender$ denote the set of code transformation methods that are available to the defender, where $M_{\defender} \subseteq M$.
The goal of the defender is to harden $\D$ into $\D^+$ to detect the vulnerability in $q^+$, produced by applying some code transformation methods in $M$ to $q$.

\begin{figure}[!t]
	\centering
	\includegraphics[width=.48\textwidth]{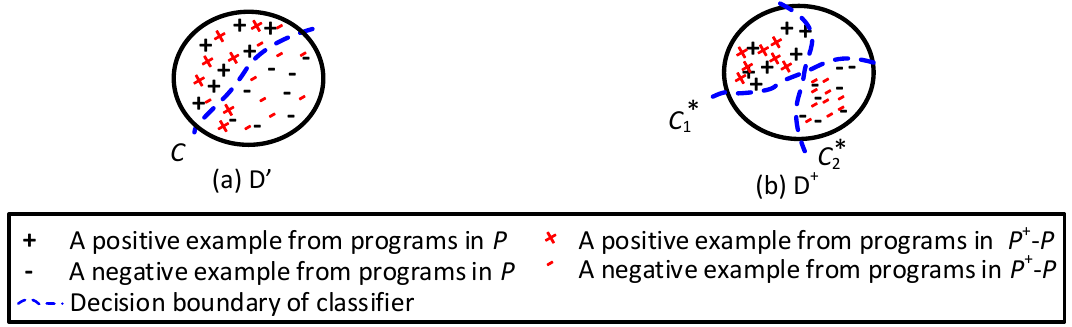}
	\vspace{-0.1cm}
	\caption{Illustrating decision boundaries of detector $\D'$ obtained via the {\em conventional} adversarial training and ZigZag-enabled detector $\D^+$}
	\label{Fig_CA_zigzag}
	\vspace{-0.6cm}
\end{figure}

\begin{figure*}[!t]
	\centering
	\subfigure[FPR]{
		\label{Fig_RQ2_FPR_slice}
		\includegraphics[width=.3\textwidth]{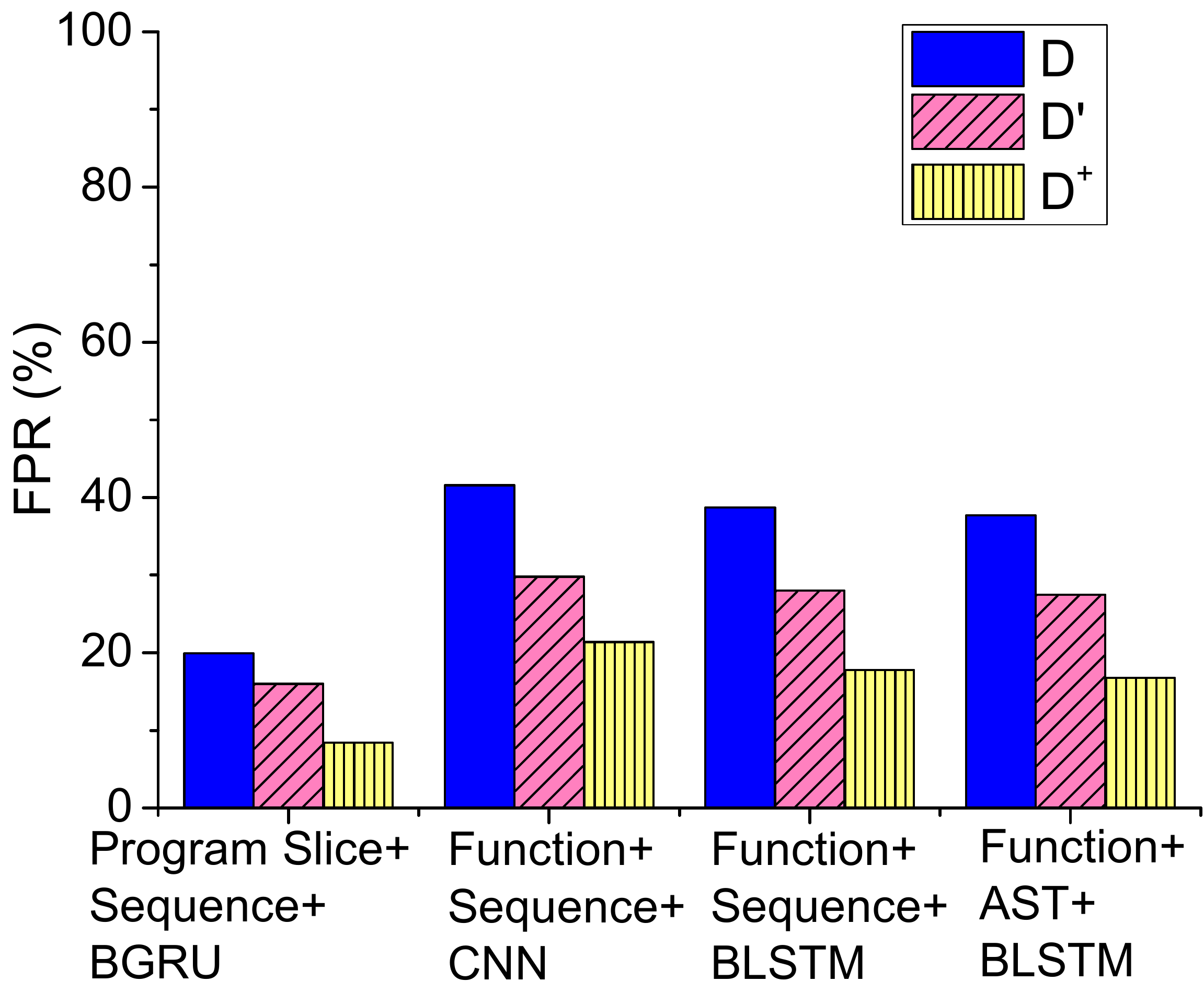}}
	\quad
	\subfigure[FNR]{
		\label{Fig_RQ2_FNR_slice}
		\includegraphics[width=.3\textwidth]{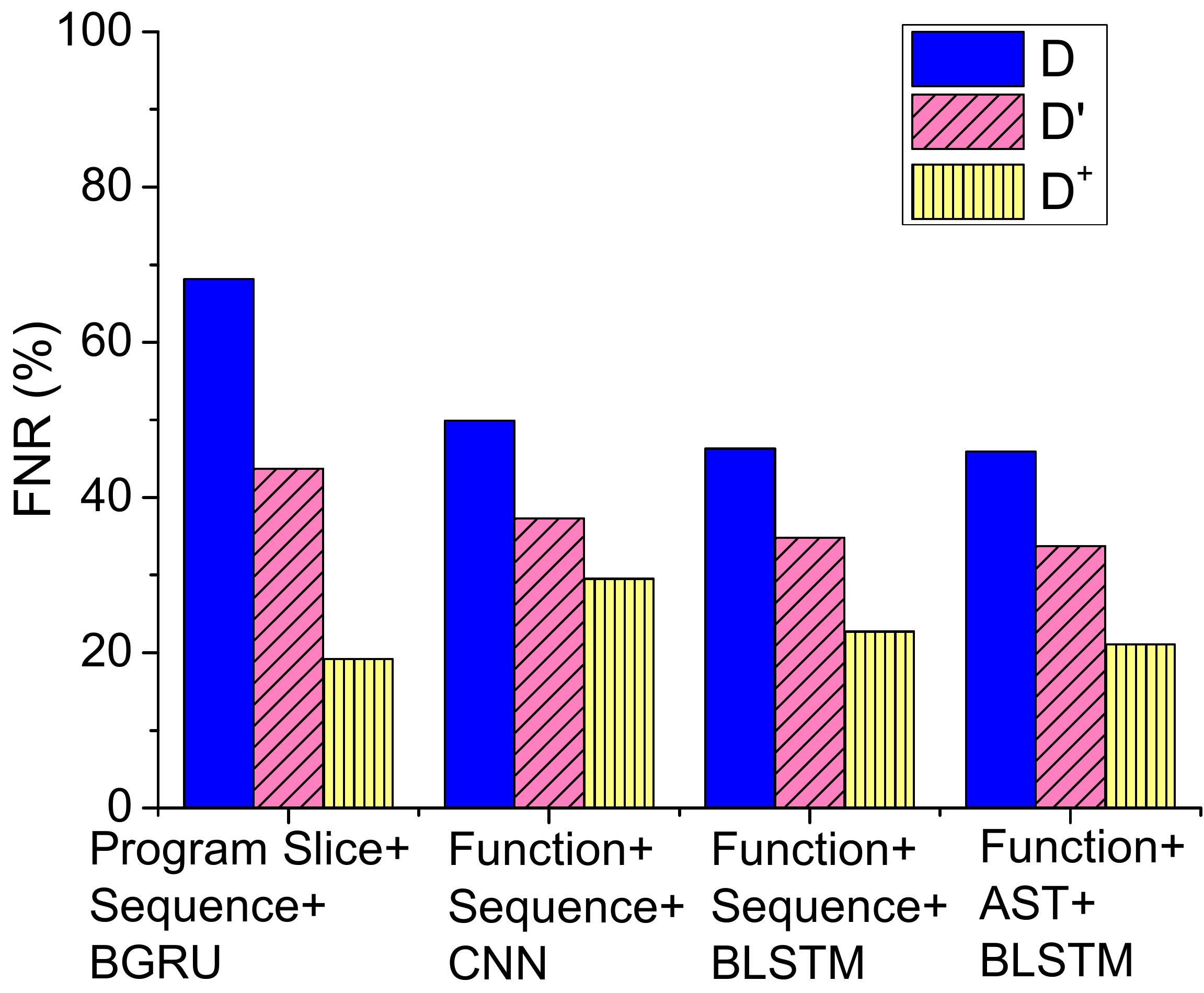}}
	\quad
	\subfigure[F1]{
		\label{Fig_RQ2_F1_slice}
		\includegraphics[width=.3\textwidth]{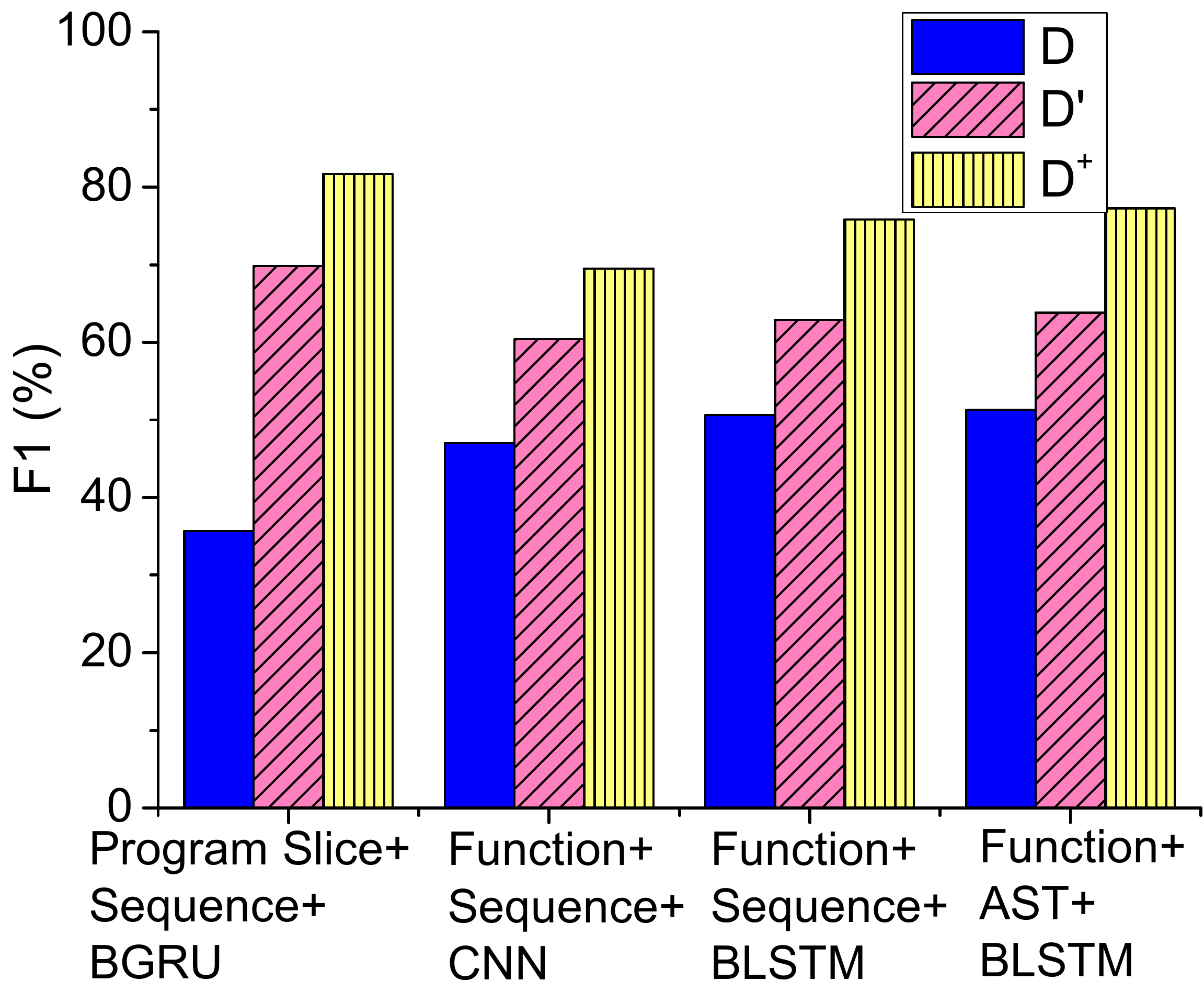}}
	\vspace{-0.3cm}
	\caption{Comparing the effectiveness of the original detector $\D$ trained from $P$, the detector $\D'$ trained from $P^+$, and the ZigZag-enabled detector $\D^+$ trained from $P^+$ with respect to four DL-based 
	detectors, where $P^+$ is derived from $P$ via CT-2, CT-7, and CT-8 in $M_{\defender,1}$.
	}
	\vspace{-0.2cm}
	\label{Fig_comparison_three_models} 
\end{figure*}

\noindent{\bf Why is conventional adversarial training incompetent?}
It may sound intuitive to use the examples corresponding to the programs in $P^+$ as input to the detector $\D$ to train a detector $\D'$, which is the {\em conventional} adversarial training. 
However, our experiments show that the effectiveness of $\D'$ is far from satisfactory 
(see Fig. \ref{Fig_comparison_three_models} in Section \ref{subsec:experiments_RQ1}).
This is because the training process tries to make the distribution of the examples corresponding to the programs in $P$ and that 
corresponding to the programs in $P^+\!\!-P$ similar, causing many examples close to the decision boundary, which would cause misclassifications with small perturbations, as shown in Fig. \ref{Fig_CA_zigzag}(a).

\noindent{\bf Basic idea}.
The ZigZag framework can be seen as a ``compiler'' that compiles an input DL-based detector, which uses a single classifier, into a {\em robust} detector, which uses two classifiers. The key insight is that adversarial vulnerability examples often reside near the boundary of a classifier. Since it is possible that there are always some examples residing near the boundary of any given classifier, the ZigZag framework leverages {\em two} classifiers with distant decision boundaries and assures that a successful adversarial example must ``fool'' these two classifiers, which is harder to achieve when the two classifiers are required to predict consistently. 
This intuition can be enforced by decoupling  {\em feature learning} and {\em classifier learning} when training a  DL-based detector as follows:
(i) feature learning aims at optimizing the feature representation such that the two classifiers use different decision boundaries but predict consistently, which implies robust features; 
and (ii) classifier learning aims at optimizing two classifiers by ``pushing'' their boundaries away from each other, as illustrated in Fig. \ref{Fig_CA_zigzag}(b) where the ZigZag-enabled detector $\D^+$ consists of classifiers $C_1^*$ and $C_2^*$ with distant decision boundaries. Putting the preceding (i) and (ii) together, when the training process converges, a ZigZag-enabled  detector is hard to evade because an adversarial example must ``fool'' both classifiers.

Fig. \ref{Fig_Overview_of_our_method} highlights the ZigZag framework. A {\em ZigZag-enabled detector} differs from a {\em DL-based detector} in the training phase, as shown in
Fig. \ref{Fig_Overview_of_our_method}(b) vs. Fig. \ref{Fig_Overview_of_our_method}(a). Note that as depicted in Figure \ref{Fig_Overview_of_our_method}(c), they share the same detection phase. 
The training of a DL-based detector has three steps, namely Steps 1, 2, and 3. For training a ZigZag-enabled detector, a new step (Step 0) is introduced 
and Step 3 is extended to what is called Step 3'.

\noindent{\bf Step 0: augmenting source code}.
The input includes (i) the source code of a set of training programs $P$ and (ii) the set of code transformations $M_\defender$ that is available to the defender. 
This step generates a set of programs, denoted by $P^+$, which includes all programs in $P$ and the programs that are transformed from the programs in $P$ via the code transformations in $M_\defender$, meaning $P\subset P^+$.
This step mimics the way a defender generates additional examples for {\em adversarial training} 
\cite{DBLP:conf/ccs/HuangJNRT11}.

\noindent{\bf Step 1: generating code fragments}. 
This is the same as in training a DL-based detector.

\noindent{\bf Step 2: mapping code fragments into vectors}. 
This is the same as in training a DL-based detector.

\noindent{\bf Step 3': training a ZigZag-enabled detector}.
The goal of this step is to learn vulnerability features robust against code transformations. 
Denote the set of code fragment vectors by $X=\{{\bf x_1}, \dots, {\bf x_\mu}\}$  for all code fragments (i.e., examples) generated from the training programs in $P$, where ${\bf x_s}$ ($1 \leq s \leq \mu$) is the vector of a code fragment with label $y_s$ ($y_s \in \{0, 1\}$ where ``0'' is non-vulnerable and ``1'' is vulnerable).
We denote the set of all code fragments generated from the manipulated programs in $P^+\!\!-P$ by $X'=\{{\bf x'_1}, \dots, {\bf x'_\nu}\}$, where ${\bf x'_w}$ ($1 \leq w \leq {\bf \nu}$) is the vector of a code fragment with label $y'_w$ ($y'_w \in \{0, 1\}$). This step has three substeps.

\begin{figure}[!t]
	\centering
	\includegraphics[width=.48\textwidth]{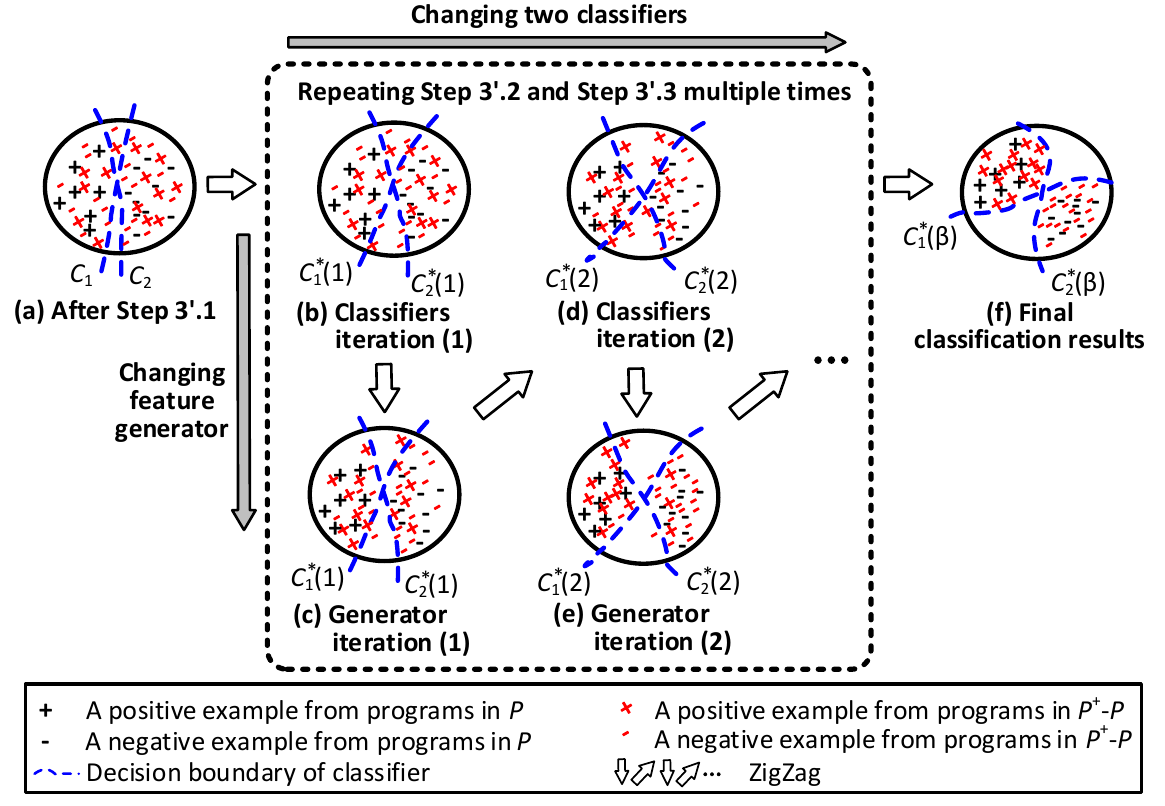}
	\vspace{-0.2cm}
\caption{The ZigZag framework iteratively tunes the classifiers and the feature generator, where a positive (negative) example indicates a vulnerable (non-vulnerable) code fragment.
} 
	\label{Fig_Step3_example}
	\vspace{-0.6cm}
\end{figure}

\smallskip
\noindent{\bf Step 3'.1: training a feature generator and two classifiers}. 
This substep aims to train a feature generator $F$ and two classifiers $C_1$ and $C_2$ to correctly classify almost all examples from programs in $P$. 
Let $c_1({\bf x_s}, F)$ and $c_2({\bf x_s}, F)$ respectively be the probability that $C_1$ and $C_2$ predict ${\bf x_s}$ as vulnerable when using $F$, where $0 \leq c_1({\bf x_s}, F), c_2({\bf x_s}, F) \leq 1$. 
Classifiers $C_1$ and $C_2$ are the same as
classifier $C$ except that they use different initial parameter values. 
We initialize $C_1$ and $C_2$ differently at the beginning of training. 
We use all examples in $X$ to train $F$, $C_1$, and $C_2$ to minimize the classification loss, which is the sum of 
cross entropies of $C_1$ and $C_2$:

\vspace{-0.6cm}
\begin{equation}
	\begin{split}
		\mathop{\min}_{F, C_1, C_2}~ -\sum_{l=1}^2 \Big \{&E_{{\bf x_s} \thicksim X} \big [y_s \cdot \log(c_l({\bf x_s}, F))+\\
		&(1-y_s) \cdot \log(1-c_l({\bf x_s}, F)) \big ] \Big \}, 
	\end{split}
\end{equation}
where $E_{{\bf x_s} \thicksim X}[\cdot]$ denotes the statistical expectation in $X$. 
Fig. \ref{Fig_Step3_example}(a) illustrates an instance of classification results of $C_1$ and $C_2$ after Step 3'.1. Almost all examples from programs in $P$ are correctly classified, while many examples from manipulated programs in $P^+\!\!\!-\!P$ are incorrectly classified.

\smallskip
\noindent{\bf Step 3'.2: training two classifiers while attempting to maximize their prediction discrepancy}. 
Because the classifiers $C_1$ and $C_2$ obtained from Step 3'.1 may be similar, 
this substep aims to train $C_1^*$ and $C_2^*$ to maximize their prediction discrepancy. 
The initial values of $C_1^*$ and $C_2^*$ are $C_1$ and $C_2$ obtained from Step 3'.1. 
For the examples in $X$, we follow the training process of Step 3'.1 to minimize the classification loss, because Step 3'.1 is able to ensure that there are hardly any incorrectly classified examples in $X$; 
for the examples in $X'$, we transform the misclassification of examples into the large prediction discrepancy of two classifiers. 

Let $c_1^*({\bf x_s}, F)$ and $c_2^*({\bf x_s}, F)$ respectively be the probability that $C_1^*$ and $C_2^*$ predict ${\bf x_s}$ as vulnerable when using feature generator $F$.
For the examples in $X$, the classification loss  
$L_c(X)$ is the sum of cross entropies of $C_1^*$ and $C_2^*$:
\vspace{-0.2cm}
\begin{equation}
	\begin{split}
		\label{equation_1}
		L_c(X) = -\sum_{l=1}^2 \Big \{&E_{{\bf x_s} \thicksim X} \big [y_s \cdot \log(c_l^*({\bf x_s}, F))+\\&(1-y_s) \cdot \log(1-c_l^*({\bf x_s}, F)) \big ] \Big \}.
	\end{split}
\end{equation}
From $X'$, we identify the examples for which $C_1^*$ or $C_2^*$ make an incorrect prediction; we call them {\em hard examples} and denote them by a set $X'' \subseteq X'$.
Hard examples satisfy (i) $C_1^*$ and $C_2^*$ make different predictions, meaning one makes a wrong prediction, or (ii) both classifiers make wrong predictions. 
We denote the set of hard examples by $X''=\{{\bf x}''_1, \dots, {\bf x}''_\gamma\}$ where ${\bf x}''_d$ ($1 \leq d \leq \gamma\;$) is the vector of a code fragment with label $y''_d \in \{0, 1\}$, which is obtained as follows:
\begin{equation}
	\label{equation_3}
	\begin{aligned}
		X''= {\sf hard\_example}\big (&X', \; {\sf binary}(c_1^*({\bf x'_w}, F)) \neq y'_w \; \vee \\ &{\sf binary}(c_2^*({\bf x'_w}, F)) \neq y'_w \big ),
	\end{aligned}
\end{equation}
where ${\sf binary}$ is a function that maps a probability to a label ``1'' or ``0'' 
and ${\sf hard\_example}$ is a function that outputs the hard examples satisfying ${\sf binary}(c_1^*({\bf x'_w}, F)) \neq y'_w$ or ${\sf binary}(c_2^*({\bf x'_w}, F)) \neq y'_w $. 
If the probability that an example ${\bf x'_w}$ is predicted as vulnerable is greater than threshold $\delta$ with $0<\delta<1$, the prediction is ``1''; otherwise, the prediction is ``0''. 
The discrepancy loss of two classifiers for hard examples $L_h(X'')$ is the absolute values of the difference between the probabilities that two classifiers predict them as vulnerable: 

\vspace{-0.4cm}
\begin{equation}
\label{equation_4}
L_h(X'')=E_{{\bf x''_d} \thicksim X''}\big [|c_1^*({\bf x''_d}, F)-c_2^*({\bf x''_d}, F)|\big ].
\end{equation}

In summary, we train $C_1^*$ and $C_2^*$ while 
using a fixed feature learner $F$ to minimize the classification loss for the examples in $X$ (i.e., minimize $L_c(X)$) and maximize the discrepancy loss of two classifiers for the examples in $X''$ (i.e., minimize $-L_h(X'')$), which can be represented as

\begin{equation}
\label{equation_5}
\mathop{{\rm min}}_{C_1^*, C_2^*}~ \big [L_c(X)-L_h(X'')\big ].
\end{equation}

Fig. \ref{Fig_Step3_example}(b) illustrates the classification of $C_1^*$ and $C_2^*$ after the first iteration in Step 3'.2. 
``Fig. \ref{Fig_Step3_example}(a)$\to$(b)'' shows the training of new classifiers, denoted by $C_1^*(1)$ and $C_2^*(1)$ where ``$(1)$'' indicates the first iteration that continues the training but correspondingly starting at $C_1$ and $C_2$ (rather than from scratch). This explains the changes in the two classification boundaries in Fig. \ref{Fig_Step3_example}(b). Note that the training of classification functions penalizes the discrepancy between the classification of $C_1^*(1)$ and that of $C_2^*(1)$.
For ``Fig. \ref{Fig_Step3_example}(a)$\to$(b)'', the positions of the examples remain unchanged and the decision boundaries are changed because the two classifiers are changed. 

\noindent{\bf Step 3'.3: training a feature generator while attempting to minimize the two classifiers' prediction discrepancy}. 
To generate robust vulnerability features, this substep aims to train the feature generator $F^*$ to learn the features that minimize the prediction discrepancy of $C_1^*$ and $C_2^*$. 
The initial value of $F^*$ is $F$ obtained from Step 3'.1. 
We use all examples in $X'$ and fixed $C_1^*$ and $C_2^*$ from Step 3'.2 to train $F^*$. The objective is to minimize the discrepancy loss of $C_1^*$ and $C_2^*$:   
\begin{equation}
\label{equation_6}
\mathop{{\rm min}}_{F^*}~ E_{{\bf x'_w} \thicksim X'}\big [|c_1^*({\bf x'_w}, F^*)-c_2^*({\bf x'_w}, F^*)|\big ].
\end{equation}

Fig. \ref{Fig_Step3_example}(c) illustrates an instance of classification results of $C_1^*$ and $C_2^*$ after the first iteration of Step 3'.3. 
Fig. \ref{Fig_Step3_example} (b)$\to$(c) corresponds to learning new features, denoted by $F^*(1)$ where ``$(1)$'' indicates the first iteration and the training continues, resuming from $F$ (rather than from scratch). Note that the classification boundaries, namely $C_1^*(1)$ and $C_2^*(1)$, remain unchanged.
For Fig. \ref{Fig_Step3_example}(b)$\to$(c), the positions of the examples are changed and the decision boundaries remain unchanged because the feature generator is changed.

The preceding Steps 3'.2 and 3'.3 are iterated, where each step guides the tuning of the other in an iterative fashion.
Fig. \ref{Fig_Step3_example} (c)$\to$(d) corresponds to training new classifiers, denoted by $C_1^*(2)$ and $C_2^*(2)$ where ``$(2)$'' indicates the second iteration, by continuing the training process but respectively starting at $C_1^*(1)$ and $C_2^*(1)$. This explains the changes in the two classification boundaries in Fig. \ref{Fig_Step3_example}(d). Note that the training of classification functions penalizes the discrepancy between the classifications of $C_1^*(2)$ and $C_2^*(2)$. 
Fig. \ref{Fig_Step3_example} (d)$\to$(e) corresponds to learning new features, denoted by $F^*(2)$ where ``$(2)$'' indicates the second iteration. Training continues, resuming from $F^*(1)$. Note that the classification boundaries, namely $C_1^*(2)$ and $C_2^*(2)$, remain unchanged. 
This ZigZag learning process proceeds by simultaneously making (i) $F^*(\beta)$ converge and (ii) 
$C_1^*(\beta)$ and $C_2^*(\beta)$ converge.  
Intuitively, this {\em joint convergence} indicates, in a sense, the identification of features and classifications that are insensitive to changes, which leads to robustness against adversarial examples.

\section{Evaluation of ZigZag-enabled Robustness}
\label{sec:experiments-and-results}
\begin{table*}[!htb]
	\caption{Effectiveness of the ZigZag-enabled detectors when using $M_{\defender,1}$ (metrics unit: \%) 
	}
	\vspace{-0.2cm}
	\label{Table_our_approach_slice}
	\scriptsize
	\centering
	\begin{tabular}{|c|p{.058\textwidth}<{\centering}|p{.058\textwidth}<{\centering}|p{.058\textwidth}<{\centering}||p{.045\textwidth}<{\centering}|p{.045\textwidth}<{\centering}|p{.045\textwidth}<{\centering}||p{.05\textwidth}<{\centering}|p{.05\textwidth}<{\centering}|p{.05\textwidth}<{\centering}||p{.04\textwidth}<{\centering}|p{.04\textwidth}<{\centering}|p{.04\textwidth}<{\centering}|}
		\hline
		& \multicolumn{3}{c||}{Program Slice + Sequence + BGRU} & \multicolumn{3}{c||}{Function + Sequence + CNN} & \multicolumn{3}{c||}{Function + Sequence + BLSTM} & \multicolumn{3}{c|}{Function + AST + BLSTM}\\
		\hline
		CT & FPR & FNR & F1 & FPR & FNR & F1 & FPR & FNR & F1 & FPR & FNR & F1 \\
		\hline
		n/a & 2.2 & 10.9 & 92.0 & 9.8 & 17.9 & 83.1 & 7.5 & 16.5 & 85.5 & 7.5 & 14.5 & 86.6\\
		\hline
		\multicolumn{13}{|c|}{Known code transformations ($M_{\defender,1}$)} \\
		\hline
		CT-2 & 6.4 & 15.7 & 85.2 & 15.2 & 30.4 & 72.3 & 19.3 & 19.5 & 76.8 & 17.1 & 18.1 & 78.9\\
		\hline
		CT-7 & 8.7 & 19.6 & 81.2 & 22.1 & 33.9 & 66.2 & 22.6 & 20.1 & 74.5 & 19.2 & 21.6 & 75.5\\
		\hline
		CT-8 & 8.9 & 19.5 & 81.2 & 23.0 & 35.7 & 64.5 & 20.2 & 23.2 & 73.9 & 19.7 & 22.4 & 74.7\\
		\hline	
		\multicolumn{13}{|c|}{Unknown code transformations ($M-M_{\defender,1}$)} \\
		\hline
		CT-1 & 9.8 & 19.1 & 80.3 & 23.5 & 25.1 & 70.9 & 21.3 & 20.9 & 74.6 & 17.9 & 18.7 & 77.8\\
		\hline
		CT-3 & 9.7 & 21.2 & 79.5 & 23.3 & 26.7 & 70.2 & 14.9 & 28.1 & 73.9 & 15.8 & 24.3 & 75.8\\
		\hline
		CT-4 & 9.4 & 21.7 & 79.5 & 32.4 & 34.5 & 63.1 & 19.5 & 29.2 & 72.3 & 19.7 & 27.1 & 73.4\\
		\hline
		CT-5 & 10.3 & 21.6 & 78.8 & 33.4 & 35.5 & 62.1 & 20.8 & 29.5 & 71.4 & 19.2 & 25.4 & 74.8\\
		\hline
		CT-6 & 9.5 & 21.7 & 79.4 & 22.6 & 31.7 & 66.8 & 19.6 & 22.5 & 74.2 & 19.7 & 21.8 & 74.6\\
		\hline
		\hline
		Total & 8.4 & 19.2 & 81.7 & 21.4 & 29.5 & 69.5 & 17.8 & 22.7 & 75.8 & 16.8 & 21.1 & 77.3\\
		\hline
	\end{tabular}
	\vspace{-0.2cm}
\end{table*}

We conduct experiments on a computer with two NVIDIA GeForce TITAN RTX GPUs and an Intel i9-9900X CPU running at 3.50GHz, and focus on answering the following three {\em Research Questions} (RQs):

\begin{itemize}
	\item {\bf RQ1}: Are ZigZag-enabled detectors robust against code transformations? (Section \ref{subsec:experiments_RQ1})
	\item {\bf RQ2}: Does the robustness of ZigZag-enabled detectors depend on the defender’s choices of code transformations?
	(Section \ref{subsec:experiments_RQ2})
	\item {\bf RQ3}: Are ZigZag-enabled detectors more effective than other widely-used vulnerability detectors?
	(Section \ref{subsec:experiments_RQ3})
\vspace{-0.4cm}
\end{itemize}

\subsection{Robustness against Code Transformation Attacks (RQ1)}
\label{subsec:experiments_RQ1}

To evaluate the robustness of ZigZag-enabled detectors against the code transformation attacks, 
we set $M_{\defender,1}$=\{CT-2, CT-7, CT-8\}, namely $M_{\defender,1} \subseteq M_\defender$, as a concrete set of code transformation methods for the defender;
we will discuss the impact of different choices of $M_{\defender,1}$ later.
The input {\em training programs} are the programs in $P$; 
the input {\em target programs} $Q^+$ are composed of the original test programs $Q$ and their manipulated programs with 8 code transformations. 
The programs in $P^+$ are generated in Step 0 and contain 26,181 vulnerable examples and 40,994 non-vulnerable examples. 
We train the four ZigZag-enabled detectors and choose the hyperparameters that lead to the highest F1. Take the ``Program Slice + Sequence + BGRU'' detector for instance. The main hyperparameters are as follows:
the batch size is 64; the number of hidden layers is 2; the dimension of hidden vectors is 900; the dropout is 0.2; the output dimension is 512; 
the learning rate is 0.002; the number of iterations 
$\beta$ for Steps 3'.2 and 3'.3 is 8; and the probability threshold $\delta$ 
is 0.4.

Table \ref{Table_our_approach_slice} summarizes the experimental results. 
Compared with four DL-based detectors, four ZigZag-enabled detectors can respectively improve the FPR, FNR, and F1 by 18.4\%, 29.4\%, and 29.9\% on average for target programs $Q^+$. This means that the ZigZag framework can 
significantly improve the robustness of DL-based detectors, especially program slice-level ones.
When compared with three ZigZag-enabled function-level detectors, the ZigZag-enabled program slice-level detector achieves a 10.3\% lower FPR, a 5.2\% lower FNR, and a 7.5\% higher F1 on average, which can be attributed to its finer granularity. 
In addition, ZigZag-enabled detectors exhibit a similar degree of effectiveness.
Taking the ``Program Slice + Sequence + BGRU'' detector for instance,
we observe that the ZigZag-enabled program slice-level detector, when applied to the manipulated programs, can significantly improve the effectiveness in terms of all three metrics. 
We also observe that the effectiveness with respect to the known code transformations is a little higher than the effectiveness with respect to the unknown code transformations, with a 1.7\% lower FPR, a 2.8\% lower FNR, and a 3.0\% F1 on average. 
This can be explained because the examples, which correspond to target programs and are generated via known code transformations, are more similar to the examples which correspond to the training programs in $P^+$ and are generated via the same code transformations in $M_{\defender,1}$.

\noindent{\bf Comparing effectiveness of $\D$, $\D'$, and $\D^+$}. 
For each of these detectors, we consider four options: ``Program Slice + Sequence + BGRU'' vs. ``Function + Sequence + CNN'' vs. ``Function + Sequence + BLSTM'' vs. ``Function + AST + BLSTM''.
Fig. \ref{Fig_comparison_three_models} compares FPR, FNR, and F1 of the original detector $\D$ trained from $P$, detector $\D'$ trained from $P^+$ (i.e., 
conventional adversarial training), and ZigZag-enabled detector $\D^+$ trained from $P^+$.
Consider the ``Program Slice + Sequence + BGRU'' instances of these detectors.
When compared with $\D'$,
$\D^+$ achieves a 6.9\% lower FPR, a 12.9\% lower FNR, and an 11.9\% higher F1 for target programs $Q^+$. When compared with $\D$, $\D'$ achieves a 3.9\% lower FPR, a 35.6\% lower FNR, and a 34.1\% higher F1. The effectiveness gained by the ZigZag framework can be understood as follows. 
$\D'$ makes the features of the programs in $P$ and their manipulated programs in $P^+\!\!-P$ similar, which causes their examples close to the decision boundary to be misclassified. 
In contrast, $\D^+$ makes the decision boundaries of two classifiers, $C_1^*$ and $C_2^*$, largely discrepant, while reducing their classification errors.

For the computational time cost of the ``Program Slice + Sequence + BGRU'' detectors, the training and the test time of $\D$ are about 2.5 hours and 5.3 hours, respectively; the training and the test time of $\D'$ are about 4.7 hours and 5.3 hours, respectively; the training and the test time of $\D^+$ are about 11.5 hours and 5.8 hours, respectively.
The test time is relatively long because of the large numbers of test examples,  
but the averaged test time cost is only 0.029 seconds per test example for $\D$ and $\D'$ and 0.032 seconds for $\D^+$. 
The costs of $\D^+$ would be justifiable by the enhanced robustness.

\noindent{\bf Effectiveness of the ZigZag framework when applied to real-world open-source software}. As discussed in Section \ref{sec:attack-experiments-results}, we can respectively manipulate 19, 8, 10, and 12 vulnerabilities in the aforementioned three real-world open-source software products 
to evade the four DL-based detectors.
Table \ref{Table_open_source_software_CVE_zigzag} lists the vulnerabilities detected in these three software products by the four ZigZag-enabled detectors.
We observe, for instance, that the ``Program  Slice  +  Sequence + BGRU'' detector detects all of the 19 vulnerabilities,
but there is still much room for improvement because there are 
12 vulnerabilities that can be transformed to  evade the detector.
Another open problem is to explain  why some code transformations can evade the detection but others cannot. This leads to:

\begin{insight}
{\em ZigZag-enabled detectors are substantially more robust than the original DL-based detectors and the detectors obtained via  conventional adversarial training.}
\end{insight}

\begin{table}[!t]
\caption{The vulnerabilities that are obtained by manipulating the ones in the three real-world software products and detected by the four ZigZag-enabled detectors 
}
	\vspace{-0.2cm}
	\label{Table_open_source_software_CVE_zigzag}
	\centering
	\tiny
	\begin{tabular}{|p{.05\textwidth}<{\centering}|c|c|p{.1\textwidth}<{\centering}|p{.07\textwidth}<{\centering}|}
		\hline
		 DL-based detector & Software product & CVE ID & Detected code transformations & Missed code transformations\\
		\hline
		{\multirow{18}{*}{\tabincell{c}{Program\\ Slice+\\Sequence+ \\BGRU}}} & FFmpeg 2.8.2 & CVE-2017-9608 & CT-3, CT-5, CT-6 & None \\
		\cline{2-5}
		& FFmpeg 2.8.2 & CVE-2018-14395 & CT-3 & CT-5, CT-6\\
		\cline{2-5}
		& FFmpeg 2.8.2 & CVE-2018-14394 & CT-3, CT-4 & CT-5\\
		\cline{2-5}
		& FFmpeg 2.8.2 & CVE-2018-1999010 & CT-1, CT-3,  CT-5, CT-6 & CT-4, CT-7\\
		\cline{2-5}
		& FFmpeg 2.8.2 & CVE-2019-12730 & CT-1, CT-2 & None \\
		\cline{2-5}
		& Wireshark 2.0.5 & CVE-2017-6467 & CT-1, CT-2 & CT-3\\
		\cline{2-5}
		& Wireshark 2.0.5 & CVE-2017-6468 & CT-1, CT-2, CT-3 & CT-5, CT-6\\
		\cline{2-5}
		& Wireshark 2.0.5 & CVE-2017-6469 & CT-2,  CT-4, CT-5, CT-8 & CT-3\\
		\cline{2-5}
		& Wireshark 2.0.5 & CVE-2017-6474 & CT-4, CT-5, CT-6 & None\\
		\cline{2-5}
		& Wireshark 2.0.5 & CVE-2017-7702 & CT-1, CT-2 & None\\
		\cline{2-5}
		& Wireshark 2.0.5 & CVE-2017-9345 & CT-3, CT-6, CT-7, CT-8 & CT-4\\
		\cline{2-5}
 	    & Wireshark 2.0.5 & CVE-2017-11410 & CT-4, CT-6, CT-7, CT-8 & None\\
		\cline{2-5}
		& Wireshark 2.0.5 & CVE-2017-11411 & CT-4, CT-7, CT-8 & CT-6\\
		\cline{2-5}
		& Wireshark 2.0.5 & CVE-2017-13767 & CT-1, CT-4 & CT-3, CT-7\\
		\cline{2-5}
		& OpenSSL 1.1.0 & CVE-2017-3730 & CT-1, CT-2, CT-6, CT-7 & None \\
		\cline{2-5}
		& OpenSSL 1.1.0 & CVE-2017-3733 & CT-1, CT-2 & CT-3, CT-4\\
		\cline{2-5}
		& OpenSSL 1.1.0 & CVE-2018-0732 & CT-3, CT-4, CT-5 & CT-6, CT-7\\
		\cline{2-5}
		& OpenSSL 1.1.0 & CVE-2019-1543 & CT-1, CT-2, CT-3, CT-4 & None\\
		\cline{2-5}
		& OpenSSL 1.1.0 & CVE-2019-1563 & CT-1, CT-2 & CT-3, CT-4 \\
		\hline
		{\multirow{8}{*}{\tabincell{c}{Function+\\Sequence+ \\CNN}}} & FFmpeg 2.8.2 & CVE-2018-9996 & CT-1  & CT-3\\
		\cline{2-5}
		& FFmpeg 2.8.2 & CVE-2018-14395 & CT-3  & CT-4\\
		\cline{2-5}
		& FFmpeg 2.8.2 & CVE-2018-1999010 & CT-1, CT-3 & CT-5\\
		\cline{2-5}
		& Wireshark 2.0.5 & CVE-2017-6467 & CT-2 & CT-3\\
		\cline{2-5}
		& Wireshark 2.0.5 & CVE-2017-6474 & CT-4, CT-5 & CT-6\\
		\cline{2-5}
		& Wireshark 2.0.5 & CVE-2017-9344 & CT-6, CT-7 & None\\
		\cline{2-5}
		& OpenSSL 1.1.0 & CVE-2017-3733 & CT-1, CT-2 & None\\
		\cline{2-5}
		& OpenSSL 1.1.0 & CVE-2018-0735 & CT-4 & CT-5\\
		\hline
		{\multirow{10}{*}{\tabincell{c}{Function+\\Sequence +\\BLSTM}}} & FFmpeg 2.8.2 & CVE-2017-9608 & CT-3, CT-5 & None\\
		\cline{2-5}
		& FFmpeg 2.8.2 & CVE-2018-14394 & CT-3 & CT-4, CT-5\\
		\cline{2-5}
		& FFmpeg 2.8.2 & CVE-2018-1999010 & CT-1, CT-3 & CT-5\\
		\cline{2-5}
		& Wireshark 2.0.5 & CVE-2017-6467 & CT-2 & CT-3\\
		\cline{2-5}
		& Wireshark 2.0.5 & CVE=2017-6474 & CT-4, CT-5 & CT-6\\
		\cline{2-5}
		& Wireshark 2.0.5 & CVE-2017-9345 & CT-5, CT-6 & CT-3\\
		\cline{2-5}
		& Wireshark 2.0.5 & CVE-2017-11411 & CT-4 & CT-5, CT-6\\
		\cline{2-5}
		& OpenSSL 1.1.0 & CVE-2017-3733 & CT-1, CT-2 & CT-3\\
		\cline{2-5}
		& OpenSSL 1.1.0 & CVE-2018-0735 & CT-3, CT-4 & CT-5\\
		\cline{2-5}
		& OpenSSL 1.1.0 & CVE-2018-0737 & CT-3 & CT-4, CT-5\\
		\hline
		{\multirow{12}{*}{\tabincell{c}{Function+\\AST+ \\BLSTM}}} & FFmpeg 2.8.2 & CVE-2017-9608 & CT-3, CT-5 & CT-6\\
		\cline{2-5}
		& FFmpeg 2.8.2 & CVE-2018-14394 & CT-3 & CT-4, CT-5\\
		\cline{2-5}
		& FFmpeg 2.8.2 & CVE-2017-9996 & CT-1, CT-3 & CT-5\\
		\cline{2-5}
		& FFmpeg 2.8.2 & CVE-2019-12730 & CT-1 & CT-2\\
		\cline{2-5}
		& Wireshark 2.0.5 & CVE-2017-6468 & CT-2 & CT-3\\
		\cline{2-5}
		& Wireshark 2.0.5 & CVE-2017-6474 & CT-4, CT-5 & CT-6\\
		\cline{2-5}
		& Wireshark 2.0.5 & CVE-2017-9345 & CT-6, CT-7, CT-8 & CT-3, CT-4\\
		\cline{2-5}
		& Wireshark 2.0.5 & CVE-2017-11410 & CT-7, CT-8 & CT-4, CT-6 \\
		\cline{2-5}
		& Wireshark 2.0.5 & CVE-2017-13766 & CT-3, CT-4 & CT-6\\
		\cline{2-5}
		& OpenSSL 1.1.0 & CVE-2017-3733 & CT-1, CT-2, CT-3 & None\\
		\cline{2-5}
		& OpenSSL 1.1.0 & CVE-2018-0734 & CT-1, CT-2 & CT-3\\
		\cline{2-5}
		& OpenSSL 1.1.0 & CVE-2019-1543 & CT-1, CT-2, CT-3 & CT-4\\
		\hline
	\end{tabular}
	\vspace{-0.6cm}
\end{table}

\subsection{Dependence on Defender's Code Transformations (RQ2)}
\label{subsec:experiments_RQ2}

\begin{figure*}[!t]
	\centering
	\subfigure[FPR]{
		\label{Fig_RQ3_FPR_slice}
		\includegraphics[width=.3\textwidth]{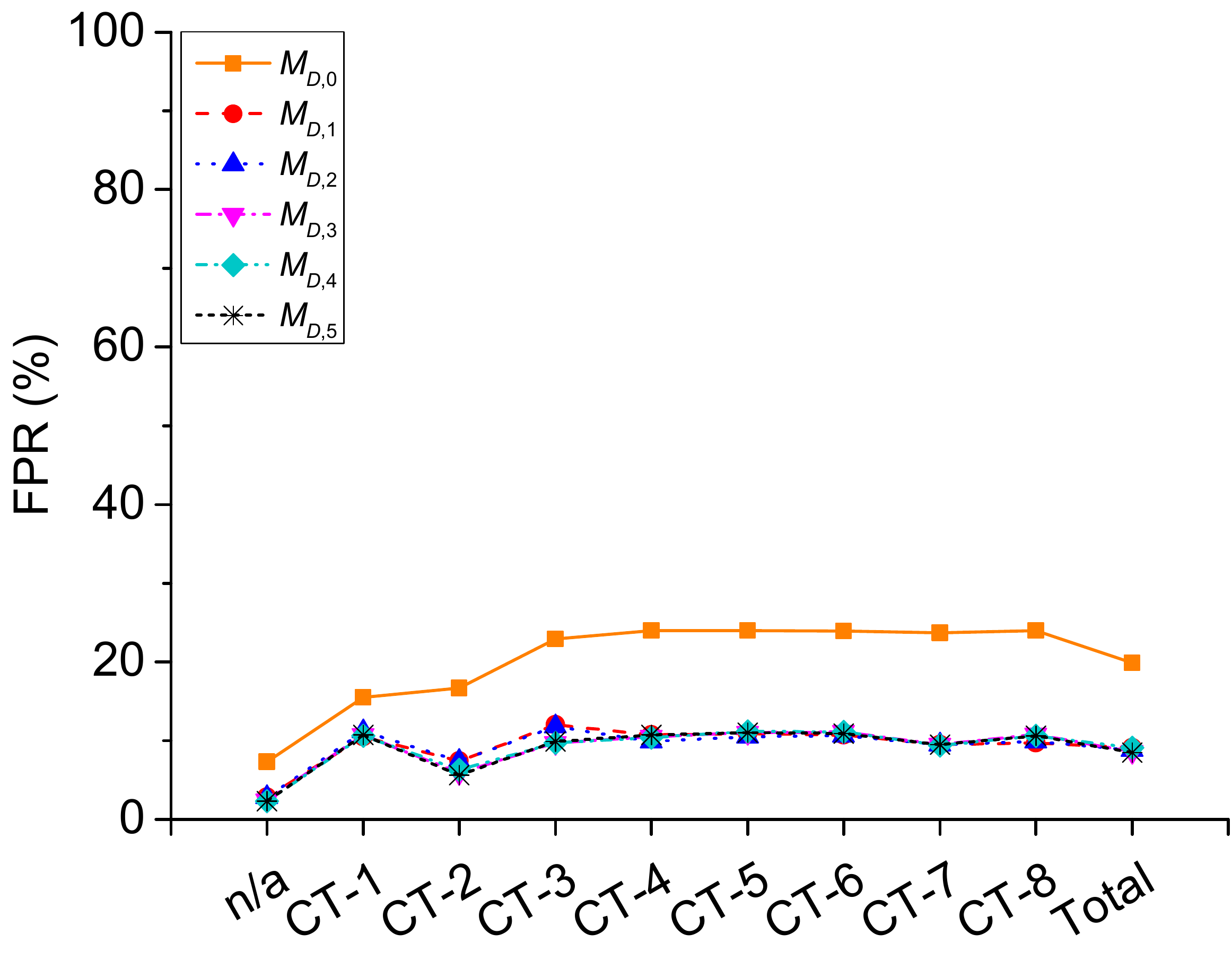}}
	\quad
	\subfigure[FNR]{
		\label{Fig_RQ3_FNR_slice}
		\includegraphics[width=.3\textwidth]{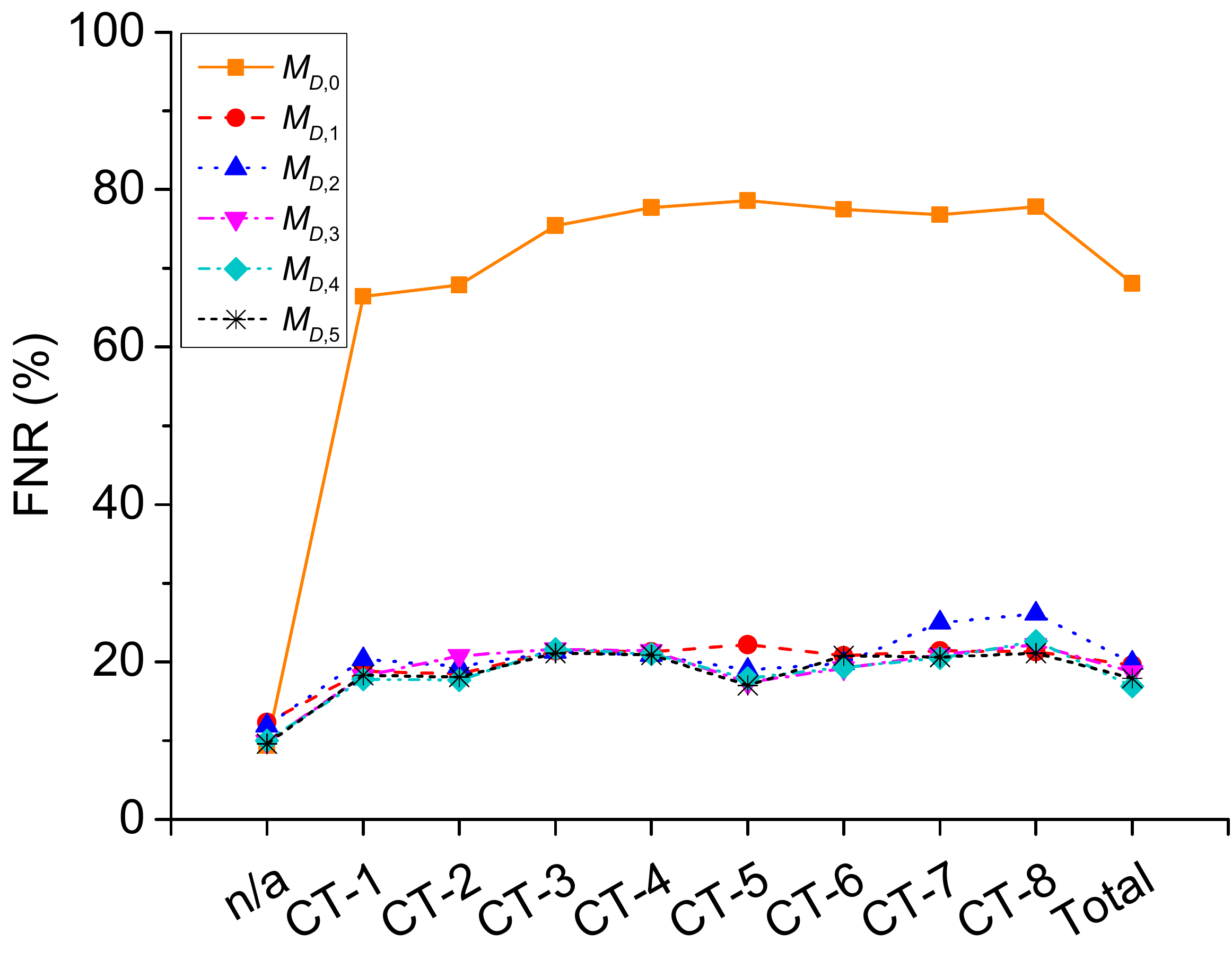}}
	\quad
	\subfigure[F1]{
		\label{Fig_RQ3_F1_slice}
		\includegraphics[width=.3\textwidth]{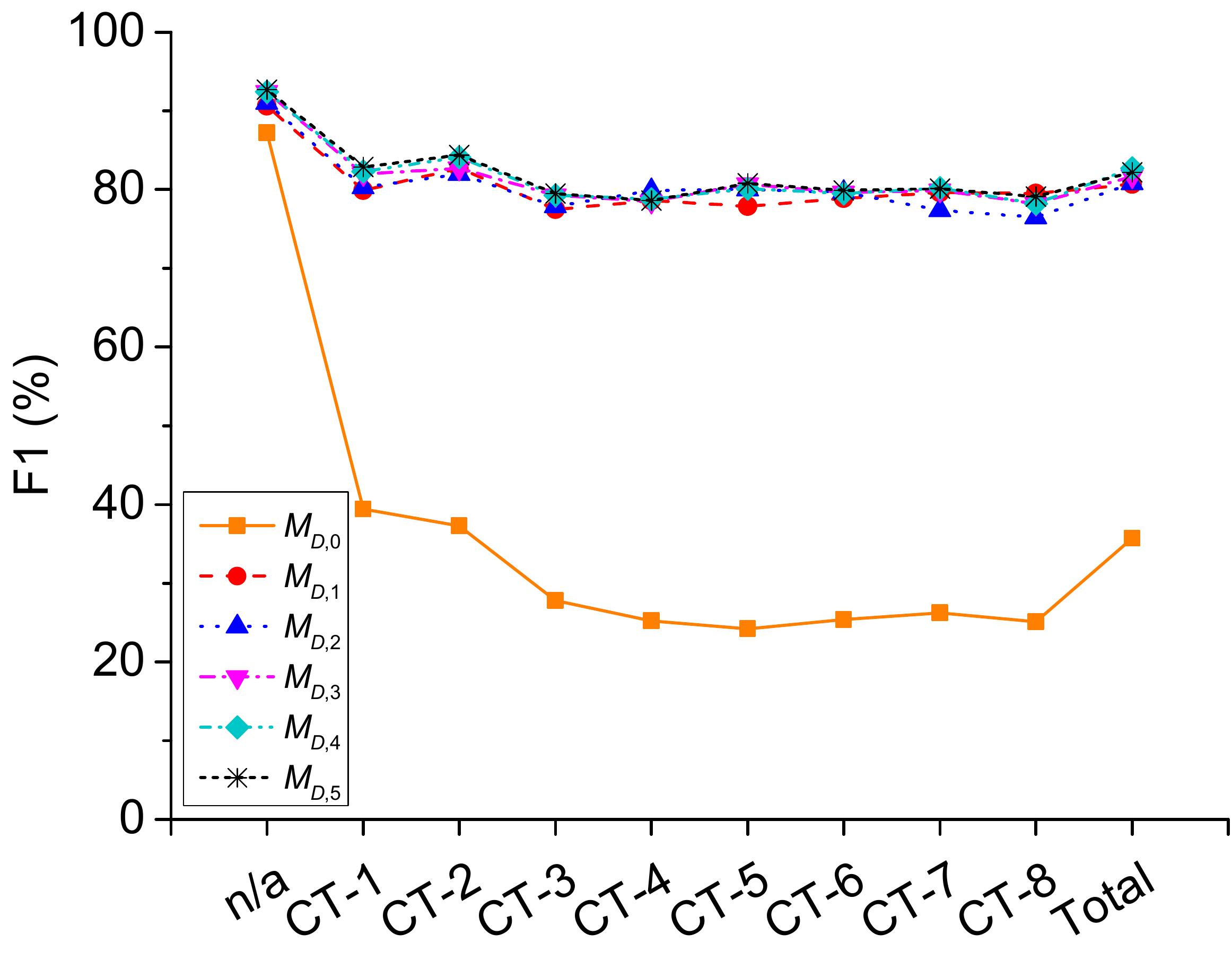}}
	\vspace{-0.3cm}
	\caption{ 
		Comparing the FPRs, FNRs, and F1s among 6 instances of $M_{\defender}$ for the ZigZag-enabled ``Program Slice + Sequence + BGRU'' }
	\vspace{-0.6cm}
	\label{Fig_Comparison_known_types_RQ3} 
\end{figure*}

To show whether the robustness of ZigZag-enabled detectors depend on the code transformations used by the defender, namely $M_{\defender}$,
we consider the following 6 instances of $M_{\defender}$: 
$M_{\defender,0}$=$\emptyset$, 
$M_{\defender,1}$=\{CT-2, CT-7, CT-8\}, $M_{\defender,2}$=\{CT-3, CT-4, CT-6\}, 
$M_{\defender,3}$=\{CT-2, CT-4, CT-5, CT-6\}, 
$M_{\defender,4}$=\{CT-1, CT-2, CT-3, CT-4, CT-6, CT-7\}, 
and $M_{\defender,5}$=$M$. 
Since the ZigZag-enabled function-level detectors achieve similar results, we consider the ZigZag-enabled ``Program Slice + Sequence + BGRU'' detector with respect to the  6 instances of $M_{\defender}$.

Fig. \ref{Fig_Comparison_known_types_RQ3} shows the comparison results. 
We observe that the ZigZag-enabled detectors using $M_{\defender,1}$ to $M_{\defender,5}$ achieve an 11.5\% lower FPR, a 47.4\% lower FNR, and a 44.1\% higher F1 on average, 
compared with using no code transformations (i.e., $M_{\defender,0}$). This indicates that introducing known code transformations during training can significantly improve the robustness of DL-based detectors.
We also observe that the ZigZag-enabled detectors using $M_{\defender,1}$ to $M_{\defender,4}$ and $M_{\defender,5}$ are very close to each other. 
Specifically, the former detectors only achieve a 0.4\% higher FPR, a 0.9\% higher FNR, and a 0.7\% lower F1 than the latter on average. 
This indicates that using several code transformations  can achieve high effectiveness.

\begin{insight}
	{\em The code transformations available to the defender
	can significantly improve the robustness of DL-based detectors against code transformations. Several known code transformations can make the ZigZag-enabled detector achieve high effectiveness for manipulated programs.}
\end{insight}

\subsection{Comparison with Other Vulnerability Detectors (RQ3)}
\label{subsec:experiments_RQ3}
We compare the effectiveness of ZigZag-enabled detectors with other widely-used vulnerability detectors. These vulnerability detectors involve a  similarity-based detector VUDDY \cite{kim2017vuddy}, an open-source rule-based tool Flawfinder \cite{FlawFinder}, a commercial rule-based tool Checkmarx \cite{Checkmarx}, a DL-based function-level detector \cite{DBLP:conf/icmla/RussellKHLHOEM18}, and a program slice-level detector SySeVR \cite{SySeVR}. 
We choose them because they are widely used to detect vulnerabilities in C programs and they are available to us.
We use $M_{\defender,1}$ 
available to the defender and use 
$Q^+$ for test.

Table \ref{Table_comparison_approaches} summarizes the experimental results. We observe that VUDDY \cite{kim2017vuddy} has a very high FNR and a low F1 because it can only detect vulnerable functions which are nearly identical to the vulnerable functions in the training programs. Therefore, most code transformations can cause VUDDY to miss vulnerabilities. Flawfinder \cite{FlawFinder} achieves a high FPR, a high FNR, and a low F1, because it does not use data flow analysis, which causes it to detect vulnerabilities inaccurately. Although  Checkmarx \cite{Checkmarx} adopts data flow analysis, its rules which are defined by human experts are far from perfect, resulting in low effectiveness. 
The DL-based function-level detector \cite{DBLP:conf/icics/LinXZX19} and SySeVR \cite{SySeVR} are effective for the original test programs, but their effectiveness drops significantly when applied to the manipulated programs. 
However, the ZigZag framework can improve their effectiveness significantly. In particular, ZigZag-enabled SySeVR achieves an 8.4\% FPR, a 19.2\% FNR, and an 81.7\% F1 when using $M_{\defender,1}$, 
which outperforms all other vulnerability detectors in our experiments.
This leads to:

\begin{insight}
	{\em The ZigZag-enabled detectors are much more effective than other kinds of vulnerability detectors when detecting vulnerabilities in manipulated programs.}
\end{insight}

\begin{table}[!t]
\caption{Comparing the effectiveness of ZigZag-enabled detectors 
and the vulnerability detectors presented in the literature} 
	\vspace{-0.2cm}
	\label{Table_comparison_approaches}
	\scriptsize
	\centering
	\begin{tabular}{|c|c|c|c|}
		\hline
		Detector & FPR (\%) & FNR (\%)& F1 (\%)\\
		\hline
		VUDDY \cite{kim2017vuddy} & 1.9  & 93.2 & 12.4   \\
		\hline
		Flawfinder \cite{FlawFinder} & 62.8  & 77.1 & 23.4 \\
		\hline
		Checkmarx \cite{Checkmarx}& 38.6  & 58.5 &  44.9 \\
		\hline
		{\tabincell{c}{DL-based function-level detector  \cite{DBLP:conf/icics/LinXZX19}}} & 38.7 & 46.3 & 50.6 \\
		\hline
		SySeVR \cite{SySeVR} & 19.9 & 68.1 & 35.7 \\
		\hline\hline
		{\tabincell{c}{ZigZag-enabled function-level detector }} & 17.8 & 22.7 & 75.8 \\
		\hline
		ZigZag-enabled SySeVR & 8.4 & 19.2 & 81.7 \\
		\hline
	\end{tabular}
	\vspace{-0.6cm}
\end{table}

\section{Limitations and Future Work}
\label{sec:Limitations}
The present study has some limitations. 
First, we focus on detecting vulnerabilities in C programs, but the methodology can be adopted or adapted to cope with other programming languages. 
Experiments need to be conducted for other languages.
Second, our experiments only consider 8 code transformations from Tigress, which are sufficient for demonstrating the feasibility of the attack and the effectiveness of the ZigZag framework. Future studies should investigate all possible code transformations.
Third, since existing datasets do not serve our purposes, the effectiveness evaluation is conducted on the dataset we collect from the NVD and SARD, which may raise an external validity issue. We have made our dataset publicly available so that third parties can repeat and validate our experiments.  
Fourth, the attack against vulnerability detectors incurs a large degree of manipulation to the source code. It is an open problem  whether or not ZigZag is effective against adversarial examples generated via small manipulations (assuming it is possible).
Fifth, the ZigZag framework uses a pair of classifiers. It is an open problem to investigate whether or not using three or more classifiers would make the resulting detectors more robust.

\section{Related Work}
\label{sec:Related_work}

\noindent{\bf Prior studies on detecting vulnerabilities.}
Our vulnerability detector leverages the static analysis of source code, which is complementary to the dynamic analysis approach \cite{manes2019art,DBLP:conf/sp/GanZQTLPC18,DBLP:conf/ccs/BohmePNR17,DBLP:conf/uss/IspoglouAMP20}.
Static analysis-based detectors can be further divided into code similarity-based, rule-based, and machine learning-based detectors. 
{\em Code similarity-based} detectors aim to detect vulnerabilities caused by code clones \cite{kim2017vuddy, DBLP:conf/acsac/LiZXJQH16, jang2012redebug}. 
{\em Rule-based} detectors use expert-defined rules to detect vulnerabilities \cite{DBLP:phd/dnb/Yamaguchi15}, including open-source tools \cite{FlawFinder,RATS,DBLP:conf/acsac/ViegaBKM00}, commercial tools \cite{Checkmarx,Coverity}, and academic efforts \cite{DBLP:conf/ndss/GensSDS18,DBLP:conf/dimva/ShastryYRS16}.
{\em Machine learning-based} detectors \cite{DBLP:conf/sigsoft/Sonnekalb19} 
aim to use models learned from expert-defined feature representations of vulnerabilities  
\cite{yamaguchi2012generalized,neuhaus2007predicting,grieco2016toward,yamaguchi2015automatic}  
or use DL models without requiring expert-defined feature representations  \cite{vuldeepecker,SySeVR,DBLP:journals/corr/abs-2001-02334,DBLP:conf/ccs/LinZLPX17,duan2019vulsniper,zhou2019devign,DBLP:journals/tii/LinZLPXVM18,DBLP:conf/ijcnn/NguyenLLNDMQP19,liu2020cd,DBLP:conf/icics/LinXZX19,DBLP:conf/sigsoft/Sonnekalb19,DBLP:conf/icmla/RussellKHLHOEM18,DBLP:journals/tifs/WangYTTHFFBW21}. 
DL-based detectors have received much attention recently. 
However, the robustness of DL-based detectors is not studied until now.

\smallskip
\noindent{\bf Prior studies on adversarial examples and training.}
Adversarial examples have attracted much attention in many domains, such as image processing \cite{goodfellow2018defense,DBLP:journals/tnn/YuanHZL19}, 
speech recognition \cite{DBLP:conf/icml/QinCCGR19}, 
malware detection \cite{li2020sok,DBLP:journals/tifs/LiL20}, program analysis \cite{rabin2019testing,DBLP:conf/aaai/Zhang20,DBLP:journals/corr/abs-1910-07517}, 
and code authorship attribution \cite{DBLP:conf/uss/QuiringMR19}. 
To the best of our knowledge, this paper is the first to study adversarial examples in the field of vulnerability detection, which make the DL-based detectors miss the vulnerabilities in the manipulated programs.
On the other hand, adversarial training is an important method to improve the robustness of DL-based models 
in the fields such as image processing \cite{DBLP:conf/iclr/MadryMSTV18,DBLP:conf/nips/ShafahiNG0DSDTG19,DBLP:conf/icse/GaoSPR20}, neural language processing \cite{DBLP:conf/ndss/LiJDLW19,DBLP:conf/icml/ZhangAD20}, malware detection \cite{DBLP:conf/uss/Chen0SJ20,li2019enhancing}, and source code processing \cite{DBLP:conf/aaai/ZhangLLMLJ20,DBLP:conf/icml/BielikV20,DBLP:journals/pacmpl/Yefet0Y20}.
The ZigZag framework proposed in this paper is an innovative adversarial training method for vulnerability detection.

\smallskip
\noindent{\bf Prior studies on code transformations.}
Code transformations are widely used for compiler-based optimizations \cite{
DBLP:journals/toplas/WhitfieldS97, DBLP:conf/ipps/KartsaklisHHIJG12, cardoso2017embedded}, program readability and maintainability improvement 
\cite{DBLP:journals/scp/KnieselK04, fowler2018refactoring,DBLP:journals/tse/DallalA18}, 
intelligent property protection \cite{DBLP:conf/icssa/KoCK17,DBLP:phd/ethos/Mirza18}, and
program analysis tasks evaluation \cite{DBLP:conf/icst/RoyC09,DBLP:journals/corr/abs-1905-11445}.
Code transformation methods are divided into three classes:
semantics-preserving vs. semantics-approximating vs. semantics-changing \cite{DBLP:journals/corr/abs-1905-11445}. 
There are some code obfuscators or program transformation tools for C programs \cite{Tigress,Stunnix,Sourceformatx,Coccinelle}. 
In this paper, we leverage semantics-preserving transformations to attack DL-based detectors.

\section{Conclusion}
\label{sec:Conclusion}
We studied the robustness of DL-based vulnerability detectors by using experiments to demonstrate that simple attacks can make vulnerabilities evade them.
We presented an innovative ZigZag framework to enhance the robustness of DL-based vulnerability detectors. 
The key insight underlying the framework is to decouple feature learning and classifier learning and make the resulting features and classifiers robust against code transformations. Experimental results show that the ZigZag framework can substantially improve the robustness of DL-based detectors.
The limitations discussed in Section \ref{sec:Limitations} offer open problems for future research.

\section*{Acknowledgment}
The authors from Huazhong University of Science and Technology were supported in part by the National Natural Science Foundation of China under Grant No. U1936211. 
S. Xu was supported in part by ARO Grant \#W911NF-17-1-0566 as well as NSF Grants \#1814825 and \#1736209. Any opinions, findings, conclusions or recommendations expressed in this work are those of the authors and do not reflect the views of the funding agencies in any sense.

\ifCLASSOPTIONcaptionsoff
  \newpage
\fi



%
\bibliographystyle{IEEEtran}
\bibliography{IEEEabrv,bibliography}

\begin{thebibliography}{10}
\providecommand{\url}[1]{#1}
\csname url@samestyle\endcsname
\providecommand{\newblock}{\relax}
\providecommand{\bibinfo}[2]{#2}
\providecommand{\BIBentrySTDinterwordspacing}{\spaceskip=0pt\relax}
\providecommand{\BIBentryALTinterwordstretchfactor}{4}
\providecommand{\BIBentryALTinterwordspacing}{\spaceskip=\fontdimen2\font plus
\BIBentryALTinterwordstretchfactor\fontdimen3\font minus
  \fontdimen4\font\relax}
\providecommand{\BIBforeignlanguage}[2]{{%
\expandafter\ifx\csname l@#1\endcsname\relax
\typeout{** WARNING: IEEEtran.bst: No hyphenation pattern has been}%
\typeout{** loaded for the language `#1'. Using the pattern for}%
\typeout{** the default language instead.}%
\else
\language=\csname l@#1\endcsname
\fi
#2}}
\providecommand{\BIBdecl}{\relax}
\BIBdecl

\bibitem{CVE}
``{Common Vulnerabilities and Exposures},'' \url{http://cve.mitre.org/}, 2020.

\bibitem{LinuxFoundation2020}
``{Open Source Software Supply Chain Security},''
  \url{https://linuxfoundation.org/wp-content/uploads/oss_supply_chain_security.pdf},
  2020.

\bibitem{Heartbleed}
Synopsys, ``The heartbleed bug,'' \url{https://heartbleed.com/}, 2014.

\bibitem{npm}
``Compromised npm package: event-stream,''
  \url{https://medium.com/intrinsic/compromised-npm-package-event-stream-d47d08605502},
  2018.

\bibitem{manes2019art}
V.~J.~M. Man{\`e}s, H.~Han, C.~Han, S.~K. Cha, M.~Egele, E.~J. Schwartz, and
  M.~Woo, ``The art, science, and engineering of fuzzing: A survey,''
  \emph{{IEEE} Trans. Software Eng.}, 2019.

\bibitem{DBLP:conf/sp/GanZQTLPC18}
S.~Gan, C.~Zhang, X.~Qin, X.~Tu, K.~Li, Z.~Pei, and Z.~Chen, ``{CollAFL}: Path
  sensitive fuzzing,'' in \emph{Proceedings of 2018 {IEEE} Symposium on
  Security and Privacy (S\&P), San Francisco, California, {USA}}, 2018, pp.
  679--696.

\bibitem{DBLP:conf/ccs/BohmePNR17}
M.~B{\"{o}}hme, V.~Pham, M.~Nguyen, and A.~Roychoudhury, ``Directed greybox
  fuzzing,'' in \emph{Proceedings of 2017 {ACM} {SIGSAC} Conference on Computer
  and Communications Security (CCS), Dallas, TX, USA}, 2017, pp. 2329--2344.

\bibitem{DBLP:conf/uss/IspoglouAMP20}
K.~K. Ispoglou, D.~Austin, V.~Mohan, and M.~Payer, ``Fuzzgen: Automatic fuzzer
  generation,'' in \emph{Proceedings of the 29th {USENIX} Security Symposium},
  2020, pp. 2271--2287.

\bibitem{kim2017vuddy}
S.~Kim, S.~Woo, H.~Lee, and H.~Oh, ``{VUDDY: A} scalable approach for
  vulnerable code clone discovery,'' in \emph{Proceedings of 2017 {IEEE}
  Symposium on Security and Privacy (S\&P), San Jose, CA, USA}, 2017, pp.
  595--614.

\bibitem{DBLP:conf/acsac/LiZXJQH16}
Z.~Li, D.~Zou, S.~Xu, H.~Jin, H.~Qi, and J.~Hu, ``{VulPecker: An} automated
  vulnerability detection system based on code similarity analysis,'' in
  \emph{Proceedings of the 32nd Annual Conference on Computer Security
  Applications (ACSAC), Los Angeles, CA, USA}, 2016, pp. 201--213.

\bibitem{jang2012redebug}
J.~Jang, A.~Agrawal, and D.~Brumley, ``{ReDeBug}: {F}inding unpatched code
  clones in entire {OS} distributions,'' in \emph{Proceedings of 2012 {IEEE}
  Symposium on Security and Privacy (S\&P), San Francisco, California, {USA}},
  2012, pp. 48--62.

\bibitem{FlawFinder}
``{Flawfinder},'' \url{http://www.dwheeler.com/flawfinder}, 2019.

\bibitem{RATS}
``{Rough Audit Tool for Security},''
  \url{https://code.google.com/archive/p/rough-auditing-tool-for-security/},
  2019.

\bibitem{DBLP:conf/acsac/ViegaBKM00}
J.~Viega, J.~T. Bloch, Y.~Kohno, and G.~McGraw, ``{ITS4:} {A} static
  vulnerability scanner for {C} and {C++} code,'' in \emph{Proceedings of the
  16th Annual Computer Security Applications Conference {(ACSAC}), New Orleans,
  Louisiana, {USA}}, 2000, pp. 257--267.

\bibitem{Checkmarx}
``Checkmarx,'' \url{https://www.checkmarx.com/}, 2019.

\bibitem{Coverity}
``Coverity,'' \url{https://scan.coverity.com/}, 2019.

\bibitem{DBLP:conf/ndss/GensSDS18}
D.~Gens, S.~Schmitt, L.~Davi, and A.~Sadeghi, ``{K-Miner}: Uncovering memory
  corruption in linux,'' in \emph{Proceedings of the 25th Annual Network and
  Distributed System Security Symposium (NDSS), San Diego, California, USA},
  2018.

\bibitem{DBLP:conf/dimva/ShastryYRS16}
B.~Shastry, F.~Yamaguchi, K.~Rieck, and J.~Seifert, ``Towards vulnerability
  discovery using staged program analysis,'' in \emph{Proceedings of the 13th
  International Conference on Detection of Intrusions and Malware, and
  Vulnerability Assessment (DIMVA), San Sebasti{\'{a}}n, Spain}, 2016, pp.
  78--97.

\bibitem{yamaguchi2012generalized}
F.~Yamaguchi, M.~Lottmann, and K.~Rieck, ``Generalized vulnerability
  extrapolation using abstract syntax trees,'' in \emph{Proceedings of the 28th
  Annual Computer Security Applications Conference (ACSAC), Orlando, FL, USA},
  2012, pp. 359--368.

\bibitem{neuhaus2007predicting}
S.~Neuhaus, T.~Zimmermann, C.~Holler, and A.~Zeller, ``Predicting vulnerable
  software components,'' in \emph{Proceedings of 2007 {ACM} Conference on
  Computer and Communications Security (CCS), Alexandria, Virginia, USA}, 2007,
  pp. 529--540.

\bibitem{grieco2016toward}
G.~Grieco, G.~L. Grinblat, L.~C. Uzal, S.~Rawat, J.~Feist, and L.~Mounier,
  ``Toward large-scale vulnerability discovery using machine learning,'' in
  \emph{Proceedings of the 6th {ACM} on Conference on Data and Application
  Security and Privacy (CODASPY), New Orleans, LA, USA}, 2016, pp. 85--96.

\bibitem{yamaguchi2015automatic}
F.~Yamaguchi, A.~Maier, H.~Gascon, and K.~Rieck, ``Automatic inference of
  search patterns for taint-style vulnerabilities,'' in \emph{Proceedings of
  2015 {IEEE} Symposium on Security and Privacy (S\&P), San Jose, CA, USA},
  2015, pp. 797--812.

\bibitem{vuldeepecker}
Z.~Li, D.~Zou, S.~Xu, X.~Ou, H.~Jin, S.~Wang, Z.~Deng, and Y.~Zhong,
  ``{VulDeePecker}: A deep learning-based system for vulnerability detection,''
  in \emph{Proceedings of the 25th Annual Network and Distributed System
  Security Symposium (NDSS), San Diego, California, USA}, 2018.

\bibitem{SySeVR}
Z.~Li, D.~Zou, S.~Xu, H.~Jin, Y.~Zhu, and Z.~Chen, ``{SySeVR}: {A} framework
  for using deep learning to detect software vulnerabilities,'' \emph{{IEEE}
  Trans. Dependable Sec. Comput.}, doi: 10.1109/TDSC.2021.3051525, 2021.

\bibitem{DBLP:journals/corr/abs-2001-02334}
D.~Zou, S.~Wang, S.~Xu, Z.~Li, and H.~Jin, ``{\(\mu\)}{VulDeePecker}: {A} deep
  learning-based system for multiclass vulnerability detection,'' \emph{{IEEE}
  Trans. Dependable Sec. Comput.}, doi: 10.1109/TDSC.2019.2942930, 2019.

\bibitem{DBLP:conf/ccs/LinZLPX17}
G.~Lin, J.~Zhang, W.~Luo, L.~Pan, and Y.~Xiang, ``{POSTER}: Vulnerability
  discovery with function representation learning from unlabeled projects,'' in
  \emph{Proceedings of 2017 {ACM} {SIGSAC} Conference on Computer and
  Communications Security (CCS), Dallas, TX, USA}, 2017, pp. 2539--2541.

\bibitem{duan2019vulsniper}
X.~Duan, J.~Wu, S.~Ji, Z.~Rui, T.~Luo, M.~Yang, and Y.~Wu, ``{VulSniper}: Focus
  your attention to shoot fine-grained vulnerabilities,'' in \emph{Proceedings
  of the 28th International Joint Conference on Artificial Intelligence
  (IJCAI), Macao, China}, 2019, pp. 4665--4671.

\bibitem{zhou2019devign}
Y.~Zhou, S.~Liu, J.~Siow, X.~Du, and Y.~Liu, ``Devign: Effective vulnerability
  identification by learning comprehensive program semantics via graph neural
  networks,'' in \emph{Proceedings of 2019 Annual Conference on Neural
  Information Processing Systems (NeurIPS), Vancouver, BC, Canada}, 2019, pp.
  10\,197--10\,207.

\bibitem{DBLP:journals/tii/LinZLPXVM18}
G.~Lin, J.~Zhang, W.~Luo, L.~Pan, Y.~Xiang, O.~Y. de~Vel, and P.~Montague,
  ``Cross-project transfer representation learning for vulnerable function
  discovery,'' \emph{{IEEE} Trans. Industrial Informatics}, vol.~14, no.~7, pp.
  3289--3297, 2018.

\bibitem{DBLP:conf/ijcnn/NguyenLLNDMQP19}
V.~Nguyen, T.~Le, T.~Le, K.~Nguyen, O.~DeVel, P.~Montague, L.~Qu, and D.~Q.
  Phung, ``Deep domain adaptation for vulnerable code function
  identification,'' in \emph{Proceedings of 2019 International Joint Conference
  on Neural Networks (IJCNN), Budapest, Hungary}, 2019, pp. 1--8.

\bibitem{liu2020cd}
S.~Liu, G.~Lin, L.~Qu, J.~Zhang, O.~De~Vel, P.~Montague, and Y.~Xiang,
  ``{CD-VulD}: Cross-domain vulnerability discovery based on deep domain
  adaptation,'' \emph{{IEEE} Trans. Dependable Sec. Comput.}, doi:
  10.1109/TDSC.2020.2984505, 2020.

\bibitem{DBLP:conf/icics/LinXZX19}
G.~Lin, W.~Xiao, J.~Zhang, and Y.~Xiang, ``Deep learning-based vulnerable
  function detection: {A} benchmark,'' in \emph{Proceedings of the 21st
  International Conference on Information and Communications Security (ICICS),
  Beijing, China}, 2019, pp. 219--232.

\bibitem{DBLP:conf/sigsoft/Sonnekalb19}
T.~Sonnekalb, ``Machine-learning supported vulnerability detection in source
  code,'' in \emph{Proceedings of the 27th {ACM} Joint Meeting on European
  Software Engineering Conference and Symposium on the Foundations of Software
  Engineering (ESEC/FSE), Tallinn, Estonia}, 2019, pp. 1180--1183.

\bibitem{DBLP:conf/icmla/RussellKHLHOEM18}
R.~L. Russell, L.~Y. Kim, L.~H. Hamilton, T.~Lazovich, J.~Harer, O.~Ozdemir,
  P.~M. Ellingwood, and M.~W. McConley, ``Automated vulnerability detection in
  source code using deep representation learning,'' in \emph{Proceedings of the
  17th {IEEE} International Conference on Machine Learning and Applications
  (ICMLA), Orlando, FL, USA}, 2018, pp. 757--762.

\bibitem{DBLP:journals/tifs/WangYTTHFFBW21}
H.~Wang, G.~Ye, Z.~Tang, S.~H. Tan, S.~Huang, D.~Fang, Y.~Feng, L.~Bian, and
  Z.~Wang, ``Combining graph-based learning with automated data collection for
  code vulnerability detection,'' \emph{{IEEE} Trans. Inf. Forensics Secur.},
  vol.~16, pp. 1943--1958, 2021.

\bibitem{goodfellow2018defense}
I.~J. Goodfellow, ``Defense against the dark arts: An overview of adversarial
  example security research and future research directions,'' \emph{CoRR}, vol.
  abs/1806.04169, 2018.

\bibitem{DBLP:journals/tnn/YuanHZL19}
X.~Yuan, P.~He, Q.~Zhu, and X.~Li, ``Adversarial examples: Attacks and defenses
  for deep learning,'' \emph{{IEEE} Trans. Neural Networks Learn. Syst.},
  vol.~30, no.~9, pp. 2805--2824, 2019.

\bibitem{DBLP:conf/icml/QinCCGR19}
Y.~Qin, N.~Carlini, G.~W. Cottrell, I.~J. Goodfellow, and C.~Raffel,
  ``Imperceptible, robust, and targeted adversarial examples for automatic
  speech recognition,'' in \emph{Proceedings of the 36th International
  Conference on Machine Learning (ICML), Long Beach, California, {USA}}, 2019,
  pp. 5231--5240.

\bibitem{li2020sok}
D.~Li, Q.~Li, Y.~Ye, and S.~Xu, ``{SoK}: Arms race in adversarial malware
  detection,'' \emph{CoRR}, vol. abs/2005.11671, 2020.

\bibitem{DBLP:journals/tifs/LiL20}
D.~Li and Q.~Li, ``Adversarial deep ensemble: Evasion attacks and defenses for
  malware detection,'' \emph{{IEEE} Trans. Inf. Forensics Secur.}, vol.~15, pp.
  3886--3900, 2020.

\bibitem{rabin2019testing}
M.~R.~I. Rabin, K.~Wang, and M.~A. Alipour, ``Testing neural program
  analyzers,'' in \emph{Proceeding of the 34th IEEE/ACM International
  Conference on Automated Software Engineering (ASE), San Diego, CA, USA},
  2019.

\bibitem{DBLP:conf/aaai/Zhang20}
H.~Zhang, Z.~Li, G.~Li, L.~Ma, Y.~Liu, and Z.~Jin, ``Generating adversarial
  examples for holding robustness of source code processing models,'' in
  \emph{Proceedings of the 34th {AAAI} Conference on Artificial Intelligence
  (AAAI), New York, NY, USA}, 2020, pp. 1169--1176.

\bibitem{DBLP:journals/corr/abs-1910-07517}
N.~Yefet, U.~Alon, and E.~Yahav, ``Adversarial examples for models of code,''
  \emph{CoRR}, vol. abs/1910.07517, 2019.

\bibitem{DBLP:conf/uss/QuiringMR19}
E.~Quiring, A.~Maier, and K.~Rieck, ``Misleading authorship attribution of
  source code using adversarial learning,'' in \emph{Proceedings of the 28th
  {USENIX} Security Symposium (USENIX Security), Santa Clara, CA, USA}, 2019,
  pp. 479--496.

\bibitem{NVD}
``{National Vulnerability Database},'' \url{https://nvd.nist.gov}, 2020.

\bibitem{SARD}
``{Software Assurance Reference Dataset},''
  \url{https://samate.nist.gov/SRD/index.php}, 2020.

\bibitem{alon2019code2vec}
U.~Alon, M.~Zilberstein, O.~Levy, and E.~Yahav, ``code2vec: Learning
  distributed representations of code,'' \emph{Proceedings of the ACM on
  Programming Languages}, vol.~3, no. POPL, pp. 40:1--40:29, 2019.

\bibitem{DBLP:conf/msr/KovalenkoBBB19}
V.~Kovalenko, E.~Bogomolov, T.~Bryksin, and A.~Bacchelli, ``{PathMiner}: A
  library for mining of path-based representations of code,'' in
  \emph{Proceedings of the 16th International Conference on Mining Software
  Repositories (MSR), Montreal, Canada}, 2019, pp. 13--17.

\bibitem{DBLP:conf/nips/HarerOLRRKC18}
J.~Harer, O.~Ozdemir, T.~Lazovich, C.~P. Reale, R.~L. Russell, L.~Y. Kim, and
  S.~P. Chin, ``Learning to repair software vulnerabilities with generative
  adversarial networks,'' in \emph{Proceedings of 2018 Annual Conference on
  Neural Information Processing Systems (NeurIPS), Montr{\'{e}}al, Canada},
  2018, pp. 7944--7954.

\bibitem{Tigress}
``Tigress,'' \url{https://tigress.wtf/}, 2020.

\bibitem{Stunnix}
``Stunnix {C/C++} obfuscator,'' \url{http://stunnix.com}, 2019.

\bibitem{Sourceformatx}
``Sourceformatx,'' \url{http://www.sourceformat.com/obfuscate-code-cpp.htm},
  2019.

\bibitem{Coccinelle}
``Coccinelle,'' \url{http://coccinelle.lip6.fr/}, 2019.

\bibitem{DBLP:journals/csur/PendletonGCX17}
M.~Pendleton, R.~Garcia{-}Lebron, J.~Cho, and S.~Xu, ``A survey on systems
  security metrics,'' \emph{{ACM} Comput. Surv.}, vol.~49, no.~4, pp.
  62:1--62:35, 2017.

\bibitem{DBLP:conf/ccs/HuangJNRT11}
L.~Huang, A.~D. Joseph, B.~Nelson, B.~I.~P. Rubinstein, and J.~D. Tygar,
  ``Adversarial machine learning,'' in \emph{Proceedings of the 4th {ACM}
  Workshop on Security and Artificial Intelligence (AISec), Chicago, IL, USA},
  2011, pp. 43--58.

\bibitem{DBLP:phd/dnb/Yamaguchi15}
F.~Yamaguchi, ``Pattern-based vulnerability discovery,'' Ph.D. dissertation,
  University of G{\"{o}}ttingen, 2015.

\bibitem{DBLP:conf/iclr/MadryMSTV18}
A.~Madry, A.~Makelov, L.~Schmidt, D.~Tsipras, and A.~Vladu, ``Towards deep
  learning models resistant to adversarial attacks,'' in \emph{Proceedings of
  the 6th International Conference on Learning Representations (ICLR),
  Vancouver, BC, Canada}, 2018.

\bibitem{DBLP:conf/nips/ShafahiNG0DSDTG19}
A.~Shafahi, M.~Najibi, A.~Ghiasi, Z.~Xu, J.~P. Dickerson, C.~Studer, L.~S.
  Davis, G.~Taylor, and T.~Goldstein, ``Adversarial training for free!'' in
  \emph{Proceedings of 2019 Annual Conference on Neural Information Processing
  Systems (NeurIPS), Vancouver, BC, Canada}, 2019, pp. 3353--3364.

\bibitem{DBLP:conf/icse/GaoSPR20}
X.~Gao, R.~K. Saha, M.~R. Prasad, and A.~Roychoudhury, ``Fuzz testing based
  data augmentation to improve robustness of deep neural networks,'' in
  \emph{Proceedings of the 42nd International Conference on Software
  Engineering (ICSE), Seoul, South Korea}, 2020, pp. 1147--1158.

\bibitem{DBLP:conf/ndss/LiJDLW19}
J.~Li, S.~Ji, T.~Du, B.~Li, and T.~Wang, ``{TextBugger}: Generating adversarial
  text against real-world applications,'' in \emph{Proceedings of the 26th
  Annual Network and Distributed System Security Symposium (NDSS), San Diego,
  California, USA}, 2019.

\bibitem{DBLP:conf/icml/ZhangAD20}
Y.~Zhang, A.~Albarghouthi, and L.~D'Antoni, ``Robustness to programmable string
  transformations via augmented abstract training,'' in \emph{Proceedings of
  the 37th International Conference on Machine Learning (ICML), Virtual Event},
  2020, pp. 11\,023--11\,032.

\bibitem{DBLP:conf/uss/Chen0SJ20}
Y.~Chen, S.~Wang, D.~She, and S.~Jana, ``On training robust {PDF} malware
  classifiers,'' in \emph{Proceedings of the 29th {USENIX} Security Symposium
  (USENIX Security)}, 2020, pp. 2343--2360.

\bibitem{li2019enhancing}
D.~Li and Q.~Li, ``Enhancing robustness of deep neural networks against
  adversarial malware samples: Principles, framework, and application to
  aics’2019 challenge,'' in \emph{Proceedings of the AAAI-19 Workshop on
  Artificial Intelligence for Cyber Security (AICS), Honolulu, Hawaii, USA},
  2019.

\bibitem{DBLP:conf/aaai/ZhangLLMLJ20}
H.~Zhang, Z.~Li, G.~Li, L.~Ma, Y.~Liu, and Z.~Jin, ``Generating adversarial
  examples for holding robustness of source code processing models,'' in
  \emph{Proceedings of the 34th {AAAI} Conference on Artificial Intelligence
  (AAAI), New York, NY, USA}, 2020, pp. 1169--1176.

\bibitem{DBLP:conf/icml/BielikV20}
P.~Bielik and M.~T. Vechev, ``Adversarial robustness for code,'' in
  \emph{Proceedings of the 37th International Conference on Machine Learning
  (ICML), Virtual Event}, 2020, pp. 896--907.

\bibitem{DBLP:journals/pacmpl/Yefet0Y20}
N.~Yefet, U.~Alon, and E.~Yahav, ``Adversarial examples for models of code,''
  \emph{Proc. {ACM} Program. Lang.}, vol.~4, no. {OOPSLA}, pp. 162:1--162:30,
  2020.

\bibitem{DBLP:journals/toplas/WhitfieldS97}
D.~Whitfield and M.~L. Soffa, ``An approach for exploring code-improving
  transformations,'' \emph{{ACM} Trans. Program. Lang. Syst.}, vol.~19, no.~6,
  pp. 1053--1084, 1997.

\bibitem{DBLP:conf/ipps/KartsaklisHHIJG12}
C.~Kartsaklis, O.~R. Hernandez, C.~Hsu, T.~Ilsche, W.~Joubert, and R.~L.
  Graham, ``{HERCULES:} {A} pattern driven code transformation system,'' in
  \emph{Proceedings of the 26th {IEEE} International Parallel and Distributed
  Processing Symposium Workshops {\&} PhD Forum, Shanghai, China}, 2012, pp.
  574--583.

\bibitem{cardoso2017embedded}
J.~M.~P. Cardoso, J.~G. de~Figueiredo~Coutinho, and P.~C. Diniz, \emph{Embedded
  Computing for High Performance: Efficient Mapping of Computations Using
  Customization, Code Transformations and Compilation}.\hskip 1em plus 0.5em
  minus 0.4em\relax Morgan Kaufmann, 2017.

\bibitem{DBLP:journals/scp/KnieselK04}
G.~Kniesel and H.~Koch, ``Static composition of refactorings,'' \emph{Sci.
  Comput. Program.}, vol.~52, pp. 9--51, 2004.

\bibitem{fowler2018refactoring}
M.~Fowler, \emph{Refactoring: Improving the Design of Existing Code}.\hskip 1em
  plus 0.5em minus 0.4em\relax Addison-Wesley Professional, 2018.

\bibitem{DBLP:journals/tse/DallalA18}
J.~A. Dallal and A.~Abdin, ``Empirical evaluation of the impact of
  object-oriented code refactoring on quality attributes: {A} systematic
  literature review,'' \emph{{IEEE} Trans. Software Eng.}, vol.~44, no.~1, pp.
  44--69, 2018.

\bibitem{DBLP:conf/icssa/KoCK17}
S.~Ko, J.~Choi, and H.~Kim, ``{COAT: Code} obfuscation tool to evaluate the
  performance of code plagiarism detection tools,'' in \emph{Proceedings of
  2017 International Conference on Software Security and Assurance (ICSSA),
  Altoona, PA, USA}, 2017, pp. 32--37.

\bibitem{DBLP:phd/ethos/Mirza18}
O.~M. Mirza, ``Style analysis for source code plagiarism detection,'' Ph.D.
  dissertation, University of Warwick, Coventry, {UK}, 2018.

\bibitem{DBLP:conf/icst/RoyC09}
C.~K. Roy and J.~R. Cordy, ``A mutation/injection-based automatic framework for
  evaluating code clone detection tools,'' in \emph{Proceedings of the 2nd
  International Conference on Software Testing Verification and Validation
  (ICST), Denver, Colorado, USA}, 2009, pp. 157--166.

\bibitem{DBLP:journals/corr/abs-1905-11445}
K.~Wang and M.~Christodorescu, ``{COSET:} {A} benchmark for evaluating neural
  program embeddings,'' \emph{CoRR}, vol. abs/1905.11445, 2019.

\end{thebibliography}

\end{document}